\documentclass[journal]{IEEEtran}
\usepackage{mathtools}
\usepackage{comment}
\usepackage{graphicx}
\usepackage{graphics} 
\usepackage{epsfig} 
\usepackage{times} 
\usepackage{amsmath} 
\usepackage{amssymb}  
\usepackage[mathscr]{eucal}
\usepackage{enumerate}
\usepackage{booktabs}
\usepackage{multirow}
\usepackage{subfigure}
\usepackage{latexsym}
\usepackage{bm}
\usepackage{dsfont}
\usepackage{multicol}
\usepackage{multirow}
\usepackage{color}
\usepackage{url}
\usepackage{algorithm}
\usepackage{algorithmicx}
\usepackage{algpseudocode}
\usepackage{comment}
\usepackage{cite}
\usepackage{url}
\newtheorem{remark}{Remark}

\newtheorem{lemma}{Lemma}

\newtheorem{definition}{Definition}

\newtheorem{theorem}{Theorem}

{\renewcommand{\arraystretch}{2.0}
	\begin{bmatrix}}%
	{\end{bmatrix}
	\renewcommand{\arraystretch}{1.0}}

\renewcommand{\baselinestretch}{1}
\def\build#1_#2^#3{\mathrel{\mathop{\kern0pt#1}\limits_{#2}^{#3}}}%

\def\build#1_#2^#3{\mathrel{\mathop{\kern0pt#1}\limits_{#2}^{#3}}}%

\makeatletter
\renewcommand*\env@matrix[1][*\c@MaxMatrixCols c]{%
	\hskip -\arraycolsep
	\let\@ifnextchar\new@ifnextchar
	\array{#1}}
\makeatother

\ifCLASSINFOpdf
\else
\fi
\begin{document}
	
	\title{Robust 
	Learning-based Predictive Control for Discrete-time Nonlinear Systems with Unknown Dynamics and State Constraints\thanks{This work has been submitted to the IEEE for possible publication. Copyright may be transferred without notice, after which this version may no longer be accessible. }}
	
	%
	\author{Xinglong~Zhang,
		~Jiahang Liu, 
 ~Xin~Xu,~\IEEEmembership{Senior member,~IEEE}, Shuyou Yu,
	 		Hong~Chen,~\IEEEmembership{Senior member,~IEEE}
		\thanks{Xinglong Zhang, Jiahang Liu,  Xin Xu are with the College of Intelligence Science and Technology, National University of Defense Technology, Changsha 410073, China. email: (zhangxinglong18@nudt.edu.cn, liujiahang1992@foxmail.com, xuxin\_mail@263.net)	
		
		Shuyou Yu, Hong Chen is with the State Key Laboratory of Automotive Simulation and
		Control and the Department of Control Science and Engineering, Jilin University
		at NanLing, Changchun 130025, China (e-mail: yushuyou@126.com, chenh@jlu.edu.cn).}
		%
		\thanks{The work was supported by the National Natural Science Foundation of China under Grant 61825305 and 62003361, China Postdoctoral Science Foundation under Grant 47680, and the National Key R$\&$D Program of China
			2018YFB1305105.}}
	
	\markboth{Journal of \LaTeX\ Class Files}
	{Shell \MakeLowercase{\textit{et al.}}: Robust 
		Robust 
		Learning-based Predictive Control for Discrete-time Nonlinear Systems with Unknown Dynamics and State Constraints}

	\IEEEtitleabstractindextext{%
		\begin{abstract}
  Robust model predictive control (MPC) is a well-known control technique for model-based control with constraints and uncertainties. In classic robust tube-based MPC approaches, an open-loop control sequence is computed via periodically solving an online nominal MPC problem, which requires prior model information and frequent access to onboard computational resources. In this paper, we propose an efficient robust MPC solution based on receding horizon reinforcement learning, called r-LPC, for unknown nonlinear systems with state constraints and disturbances. The proposed r-LPC utilizes a Koopman operator-based prediction model obtained off-line from pre-collected input-output datasets.
  Unlike classic tube-based MPC, in each prediction time interval of r-LPC, we use an actor-critic structure to learn a near-optimal feedback control policy rather than a control sequence. The resulting closed-loop control policy can be learned off-line and deployed online or learned online in an asynchronous way. In the latter case, online learning can be activated whenever necessary; for instance, the safety constraint is violated with the deployed policy.  The closed-loop recursive feasibility, robustness, and asymptotic stability are proven under function approximation errors of the actor-critic networks. Simulation and experimental results on two nonlinear systems with unknown dynamics and disturbances have demonstrated that our approach has better or comparable performance when compared with tube-based MPC and LQR, and outperforms a recently developed actor-critic learning approach. %

		\end{abstract}
		
		\begin{IEEEkeywords}
			Model predictive control, reinforcement learning, state constraints, robustness, nonlinear systems.
	\end{IEEEkeywords}}

\maketitle

\IEEEdisplaynontitleabstractindextext

%
\IEEEpeerreviewmaketitle

\section{Introduction}
Model predictive control (MPC) has received considerable attention due to its theoretical developments and wide-spreading applications in industrial plants, robots, etc., see~\cite{mayne2000constrained,qin2003survey,wabersich2018linear}. 
Most of the existing MPC approaches are model-based, whose implementations rely on the knowledge of dynamics that are typically identified a priori. For control of systems with modeling errors caused by identification and possible exogenous disturbances, 
robust MPC such as min-max MPC in~\cite{bemporad2003min,bemporad1999robust} or tube-based MPC in~\cite{mayne2005robust,falugi2013getting} and the references therein, can be used to ensure robustness and constraint satisfaction. In classic tube-based MPC (also termed as tube MPC) for linear systems, the resulting control is composed of a nominal control action computed by a standard MPC and an off-line linear feedback control policy. However, in tube MPC for nonlinear systems (cf.~\cite{falugi2013getting}), two nonlinear optimization problems are typically solved, likely with an intensive computational load. Resorting to the recent developments in the machine learning community, in~\cite{lucia2018deep,paulson2020approximate,hertneck2018learning} and the references therein, deep neural networks have been used to approximate an explicit control policy of robust MPC. The control policy is learned off-line and deployed online. As a consequence,  the computational load can be dramatically reduced; however new issues such as insufficient generalization and transfer abilities might be inherited through deep neural networks.

As a class of methods for solving optimal control problems, reinforcement learning (RL) and adaptive dynamic programming (ADP) have also received significant attention in the past decades, see~\cite{wangadp2017,lewis2009reinforcement,ni2014grdhp,lewis2013reinforcement,wang2009adaptive,liu-2015,luo2017output} and the references therein. Among the classic RL and ADP approaches,  infinite-horizon optimal control problems with continuous state-space can be solved in a forward-in-time way via actor-critic learning.  Along this direction, various notable algorithms have been studied in~\cite{zhang2008novel,liu2013policy,zhang2009neural} for discrete-time nonlinear systems with prior dynamical knowledge, in~\cite{wang2012optimal,liu2012neural} for unknown discrete-time nonlinear dynamics, and in~\cite{jiang2012computational} for unknown continuous-time nonlinear dynamics.  {\color{black}The extensions to learning-based event-triggered control approaches can be found, for instance in~\cite{li2021event} for discrete-time nonlinear systems and in~\cite{9346029} for control of autonomous vehicles under denial-of-service attacks.} To solve finite-horizon optimal control problems,  a finite-horizon ADP solution was proposed in \cite{Xu-finitehorizonadp} for nonlinear discrete-time systems with input constraints, and the stability is guaranteed under an open-loop stability assumption. In~\cite{zhao2014neural}, a finite-horizon near-optimal control algorithm was presented for discrete-time nonlinear affine systems with unknown dynamics by using an identifier-actor-critic structure.
	
Due to the common roots in optimal control, the relations between MPC and RL and their comparisons have been studied in~\cite{GORGES20174920} and~\cite{mpcrl_com}, and some initial research works on the integration of MPC and RL have emerged recently, which can be called learning-based predictive control (LPC). In \cite{zanon2020safe} and~\cite{koller2019learningbased},  safe robust MPC algorithms were proposed for learning-based control with an event-triggered mechanism. In~\cite{wabersich2018linear}, a safe learning controller was presented based on set-membership recursion. From a different perspective that this paper focuses on, RL with an actor-critic structure has been used to solve learning-based predictive control problems, see e.g.,~\cite{xu2018learning,dong2018functional}. The optimization problem in each prediction interval is solved in a forward-in-time manner via actor-critic learning. In specific, a batch-mode RL-based predictive controller for discrete-time systems with stochastic noises and control constraints was proposed in~\cite{xu2018learning}. In~\cite{dong2018functional}, an ADP-based functional MPC was proposed for nonlinear discrete-time systems, where the control saturation is used to deal with the control constraint. The uniform ultimate boundedness of the closed-loop system is proven under certain conditions, relying on a stability assumption similar to~\cite{Xu-finitehorizonadp}. 

In the framework of LPC, the closed-loop robustness for unstable (stabilizable) perturbed systems under actor-critic approximation errors is a critical issue, which has not been addressed in~\cite{Xu-finitehorizonadp,zhao2014neural,xu2018learning,dong2018functional}. Also, the trial-and-error learning manner of RL could lead to a state constraint violation, which is expensive for safety-critical systems. To the best of our knowledge, no prior LPC with actor-critic network has addressed this point. These issues motivated our work.

In this paper, we propose a robust learning-based predictive controller using actor-critic structure, i.e. r-LPC, for a class of discrete-time nonlinear systems with unknown dynamics, state constraints, and disturbances. This work can be regarded as a new development of previous robust tube-based MPC by using an RL-based strategy to learn closed-loop control policies in an efficient way. 
 The main features of the proposed r-LPC are summarized as follows.
 First, the proposed approach can learn an explicit closed-loop control policy on the state variable. As a consequence, the control policy can be learned off-line and deployed online, or learned online in an asynchronous way whenever necessary, for instance, the safety constraint is violated,  such that the online computational load can be reduced.  The closed-loop recursive feasibility, robustness, and asymptotic stability of r-LPC are proven under modeling and function approximation errors. Simulation and experimental results on two nonlinear plants with unknown dynamics and disturbances have demonstrated that our approach can obtain better or comparable performance when compared with  the previous tube-based  MPC~\cite{XLZ2021} and LQR. In addition, the proposed r-LPC approach also outperforms a recently developed actor-critic learning approach~\cite{li2020actor} in the adopted tests. 

 The rest of the paper is organized as follows.  Section II introduces the control problem and preliminaries. In Section III the theoretical property of r-LPC is given. Section VI shows the simulation and experimental results, while some conclusions are drawn in Section VII. Some Proofs and off-line computational details are given in the appendix.

 \textbf{Notation:} We denote $\mathbb{N}_{l_1}^{l_2}$ as the set of integers $l_1,l_1+1,\cdots,l_2$. Given the variable $r$, we use $\bm {r}_{k:k+N-1}$ to denote the sequence $r (k)\ldots r (k+N-1)$, where $k,N\in\mathbb{N}_0^{\infty}$. For a vector $x\in\mathbb{R}^{n}$, we denote $\|x\|_Q^2$ as $x^{\top}Qx$ and $\|x\|$  as the Euclidean norm (as the Frobenius norm if $x$ is a matrix). For a matrix $A\in\mathbb{R}^{m\times n}$, we denote $\sigma_{\rm min}(A)$ as the minimal singular value.  Given two sets $\mathcal{Z}$ and $\mathcal{V}$, their Minkowski sum  is  $\mathcal{Z}\oplus \mathcal{V}=\{z+v|z\in \mathcal{Z}, v\in \mathcal{V}\}$, and we denote $\rm Int(\mathcal{Z})$ as the interior of $\mathcal{Z}$. 
 For variables $z_{i}\in{\mathbb{R}}^{q_{i}}$,
 $i\in\mathbb{N}_1^M$, we define $(z_{1}, z_{2}, \cdots, z_{\rm\scriptscriptstyle M})=[\,z_{1}^{\top}\ z_{2}^{\top}\ \cdots\ z_{\rm\scriptscriptstyle M}^{\top}\,]^{\top}\in{\mathbb{R}}^{q}$, where $q= \sum_{i=1}^{M}q_{i}$. 	
\section{Problem formulation and preliminaries}
\subsection{Control problem and preliminary solution}\label{sec:model-mpc}
Consider a class of discrete-time nonlinear systems with additive disturbances described by
\begin{equation}\label{Eqn:non-model}
x(k+1)=f(x(k),u(k))+w(k),
\end{equation}
where $x\in\mathcal{X}\subseteq\mathbb{R}^{n}$, $u\in\mathcal{U}\subseteq\mathbb{R}^m$ are the state and control variables,
 $w\in\mathcal{W}\subset\mathbb{R}^{n}$ is a bounded additive disturbance, $\mathcal{X}$, $\mathcal{U}$, and $\mathcal{W}$ are convex sets containing the origin in their interiors, mapping $f$ can be unknown. 
It is assumed that  $f$ is $C^1$ and $f(0,0)=0$, $f<+\infty$ in the domain $\mathcal{X}\times \mathcal{U}$, the state $x$ is measurable. 

Starting from any $x(0)\in\mathcal{X}$, the control goal of interest is to minimize a cost function
$V_{\infty}\big({x}(0)\big)=
\sum_{k=0}^{+\infty}\| {x}(k)\|_{Q}^2+\| u(k)\|_{R}^2$ subject to constraints $x(k)\in \mathcal{X}$, $u(k)\in \mathcal{U}$,
where $Q=Q^{\top}\in\mathbb{R}^{n\times n}$ and $R=R^{\top}\in\mathbb{R}^{m\times m}$, $Q,R\succ0$. 


In the following, we review a recently developed tube-based Koopman MPC in~\cite{XLZ2021} for solving the considered problem.  

{\color{black} We first introduce the Koopman model based on the Koopman operator, cf.~\cite{arbabi2018introduction,korda2018linear,XLZ2021}.  To this end, let  $\phi(x):\mathcal{X}\rightarrow \mathbb{C}$ be an observable of state $x$ and  $\mathcal{F}$ be a given space of observables.
The Koopman operator describes dynamics $x(k+1)=f(x(k))$ using a linear dynamic evolution of the observable (cf.~\cite{arbabi2018introduction,korda2018linear,XLZ2021}), i.e., \begin{equation}\label{Eqn:koopman-auto}
\phi(x(k))=\mathcal{K} \phi(x(k-1))=\mathcal{K}^k \phi(x(0)),
\end{equation}
for every $\phi(x)\in\mathcal{F}$. 

To apply the Koopman operator for systems with controls, i.e.,~\eqref{Eqn:non-model}, we define an extended state space $\mathcal{X}\times \ell(\mathcal{U})$, where $\ell(\mathcal{U})$ is the space of $\{{u}_{w}(i)\}_{i=0}^{\infty}$ with $u_w(i)=(u(i),w(i))\in\mathcal{U}\times \mathcal{W}$. Letting $f_W(x,u_w):=f(x,u)+w$ and $\boldsymbol{u}_w(i)$ be the $i$-th element of $\boldsymbol{u}_w:=\{{u}_{w}(i)\}_{i=0}^{\infty}$, 
one can write the evolution of the extended state   $ \boldsymbol s=(x,\boldsymbol{u}_w)$ 
as 
\begin{equation}\label{Eqn:extended-model}
\boldsymbol s(k+1)=F(\boldsymbol s(k)),
\end{equation} 
where $F(\boldsymbol s)=(f_W(x,\boldsymbol{u}_w(0)), {\varGamma} \boldsymbol{u}_w)$, $\varGamma$ is a left shift operator such that $\boldsymbol{u}_w(i+1)=\varGamma\boldsymbol{u}_w(i)$.  Hence, one can apply the Koopman operator in~\eqref{Eqn:koopman-auto} to~\eqref{Eqn:extended-model}, (see also~\cite{arbabi2018introduction,korda2018linear,XLZ2021}), i.e.,
\begin{equation}\label{Eqn:koopman_ex}
\phi(\boldsymbol s(k))=\mathcal{K} \phi(\boldsymbol s(k-1))=\mathcal{K}^k \phi(\boldsymbol s(0))
\end{equation}
for every $\phi({\boldsymbol s}):\mathcal{X}\times\ell(\mathcal{U\times W})\rightarrow \mathbb{C}$ belonging in $\mathcal{F}_e$ which is a given space of  $\phi({\boldsymbol s})$.
  
Note that in~\eqref{Eqn:koopman_ex} the Koopman operator $\mathcal{K}$ is of infinite dimension. 
For practical concerns of controller design, a finite-dimensional approximation of $\mathcal{K}$ is required. Specially, we choose a collection of obseravles as
$\Phi(\bm s)=
(\Psi(x),u_w)$
where  $\Psi(x)=( {\psi_{1}\left(x\right)},{\cdots}, {\psi_{\bar n}\left(x\right)})$, $\bar n>n$, and $\psi_i$, $i\in\mathbb{N}_1^{\bar n}$ are constructed as basis functions. Let the approximation of Koopman operator be $\mathcal{K}_{\bar N}\in\mathbb{R}^{\bar N\times \bar N}$ associated with $\Phi(\bm s)$,  $\bar N=\bar n+m+n$. From~\eqref{Eqn:koopman_ex}, one writes 
\begin{equation}\label{Eqn:koopman-error}
\Phi(\bm s(k+1))=\mathcal{K}_{\bar N} \Phi(\bm s(k))+\varepsilon(k),
\end{equation} 
where $\varepsilon$ is the error  due to the approximation of $\mathcal{K}$.

As shown in~\cite{XLZ2021}, there exists an inverse mapping, denoted as $\Psi^{-1}$, such that $\Psi^{-1}(\Psi(x))=x$, i.e. the evolution of $x$ can be recovered with~\eqref{Eqn:koopman-auto} through $\Psi^{-1}$.}
	Let $C\Psi(x)$, $C\in\mathbb{R}^{n\times \bar n}$, be an approximation of $\Psi^{-1}$, such that $x=\Psi^{-1}(\Psi(x))=C\Psi(x)+v$ (cf.~\cite{XLZ2021}), where  the approximation error $v\in \mathcal{V}$,  $\mathcal{V}$ is a compact set containing the origin. As the mapping from $\Phi(s)$ to $\Psi(x)$ is of interest, we denote the first $\bar n$ rows of $\mathcal{K}_{\bar N}$ as $[\mathcal{K}_{\bar N}]_{1:\bar n}=[A\ B\ D]$, $A\in\mathbb{R}^{\bar n\times\bar n}$, $B\in\mathbb{R}^{\bar n\times m}$, and $D\in\mathbb{R}^{\bar n\times n}$.  Let $z=\Psi(x)$, in view of~\eqref{Eqn:koopman-error}, then one can write an equivalent form of~\eqref{Eqn:non-model} considering the modeling errors, i.e.,
\begin{equation}\label{Eqn:linear_p-residual}
\left\{\begin{array}{ll}
{z}(k+1)=A{z}(k)+Bu(k)+d(k),\ {z}(k)=\Psi\left(x(k)\right) \\
x(k)=C{z}(k)+v(k),
\end{array}\right.
\end{equation}
where $d=Dw+\bar w\in\mathcal{D}$ is the exogenous input, $\bar w=\varepsilon_{[1:\bar n]}$, $\mathcal{D}$ is a compact set containing the origin. The model parameters $A$, $B$, $C$, $D$ and sets $\mathcal{D}$, $\mathcal{V}$ are computed in a data-driven way, which are deferred in Appendix~\ref{SEC:COM-SETS}.

The following assumptions about model~\eqref{Eqn:linear_p-residual} hold:
	\begin{enumerate}
		\item[({A1})]\label{assu:ab} the matrix $A$ is stabilizable and full rank;
		\item[({A2})]\label{assu:phi-ori} the lifted function $\Psi(x)$ satisfies $\Psi(0)=0$; 
		\item[({A3})]\label{assu:phi-lip}  $\Psi(x)$ is Lipschitz continuous for all $x\in\mathcal{X}$. \hfill $\blacktriangleleft$
	\end{enumerate}
To control~\eqref{Eqn:linear_p-residual}, i.e.~\eqref{Eqn:non-model}, a robust tube MPC can be used. The overall control law is given as
\begin{equation}\label{Eqn:real_u}
u=\hat u^{\ast}+Ke_{\scriptscriptstyle{z}},
\end{equation}
where $e_z=z-\hat z$, matrix $K\in\mathbb{R}^{m\times \bar n}$ is  such that $F=A+BK$ is Schur stable, $\hat u^{\ast}$ is computed by a nominal MPC (deferred in~\eqref{Eqn:optimizL}) using the following predictor:
\begin{equation}\label{Eqn:unpert}
\left\{\begin{array}{l}
\hat{z}(k+1)=A\hat{z}(k)+B\hat u(k)\\
[0.2cm] \hat x(k)=C\hat z(k)
\end{array}\right. \qquad
\end{equation} 
By subtracting~\eqref{Eqn:linear_p-residual} with~\eqref{Eqn:real_u} and~\eqref{Eqn:unpert}, the error $e_{\scriptscriptstyle {z}}$ evolves in the following unforced system:
\begin{equation}\label{Eqn:e}
\left\{\begin{array}{l}
e_{\scriptscriptstyle{z}}(k+1)=Fe_{\scriptscriptstyle{z}}(k)+d(k)\\
[0.2cm] e_{\scriptscriptstyle{x}}(k)=C e_{\scriptscriptstyle{z}}(k)+v(k),
\end{array}\right. \qquad
\end{equation}
where $e_{\scriptscriptstyle{x}}=x-\hat x$.
Let the robust invariant set  of $e_{\scriptscriptstyle{z}}$ be $\mathcal{O}_z$, such that $ F\mathcal{O}_z\oplus\mathcal{D}\subseteq\mathcal{O}_z$.  The corresponding  robust ``output" invariant set of $e_x$ is defined as $\mathcal{O}_x=C\mathcal{O}_z\oplus \mathcal{V}$. 

Inline with~\cite{XLZ2021}, at any time $k$, a nominal MPC is solved online to compute $\hat u^{\ast}$ in~\eqref{Eqn:real_u}, i.e.,
\begin{equation}\label{Eqn:optimizL}
\begin{array}{ll}
\min_{\bm {\hat u}_{k:k+N-1}}  V=&\sum_{i=0}^{N-1}(\|\hat {z}(k+i)\|_{\bar Q}^2+\|\hat u(k+i)\|_{R}^2)+\\
&V_f(\hat{z}(k+N)),
\end{array}
\end{equation}
where $\bar Q=\bar Q^{\top}\in\mathbb{R}^{\bar n\times \bar n}$, $\bar Q\succ 0$,
$V_f(\hat{z})=\|\hat{z}\|_P^2$, $P=P^{\top}\in\mathbb{R}^{\bar n\times \bar n}$, $P\succ 0$ is the solution to
\begin{equation}\label{Eqn:LYA}
F^{\top}PF-P=-\bar Q-K^{\top}RK.
\end{equation}
 Problem~\eqref{Eqn:optimizL} is performed subject to constraints~\eqref{Eqn:unpert},  $\hat {z}(k+i)\in  \mathcal{Z}$, $\hat u(k+i)\in \hat {\mathcal{U}},\,\,i\in\mathbb{N}_{0}^{n-1}$, and $\hat {z}(k+N)\in {{\mathcal{Z}}}_{f}$,  where $\mathcal{Z}=\{\hat{z}|C\hat {z}\oplus\mathcal{O}_x\in  {\mathcal{X}}\}$,  $\hat {\mathcal{U}}\oplus K\mathcal{O}_z={\mathcal{U}}$,
${\mathcal{Z}}_{f}$ is a positive invariant set of~\eqref{Eqn:unpert} under constraints $\hat z\in\mathcal{Z}$ and $\hat u\in\hat {\mathcal{U}}$. Like~\cite{XLZ2021}, it is assumed that the sets  $\mathcal{Z}$ and $\hat {\mathcal{U}}$ are non-empty and contain the origin in the interior; otherwise, problem~\eqref{Eqn:optimizL} is infeasible.
\begin{remark}
{\color{black}	We now summarize the roles of the two control terms in the control law (7) as follows. The first term  is computed by solving a nominal MPC (see~\eqref{Eqn:optimizL}) to generate the center trajectory of the tube, where tightened constraints $\hat z\in\mathcal{Z}$ and $\hat u\in\hat {\mathcal{U}}$ are enforced  for real constraint fulfillment. The auxiliary feedback term $Ke_z$ is introduced to steer the real state $z$ to $\hat z$. } \hfill $\blacktriangle$	
\end{remark}

Let $\bm {\hat u}^{\ast}_{k+N-1|k}$ be the optimal solution to~\eqref{Eqn:optimizL} at time $k$, then the control
\begin{equation}\label{Eqn:con-applied}
u(k)=\hat u^{\ast}(k|k)+Ke_{\scriptscriptstyle{z}}(k),
\end{equation}
 is applied at time $k$, and problem~\eqref{Eqn:optimizL} is repeatedly solved according to the receding horizon principle. 
\subsection{Definitions about barrier functions and feasible control}\label{sec:nece-defin}
We also introduce a type of barrier functions about constraints to be used in the actor-critic learning algorithm for state constraint satisfaction.
\begin{definition}[Barrier function]\label{defi:polytopic}
	For any convex set $\mathcal{Z}=\{z\in\mathbb{R}^n| g_i(z)\leq 1,\forall i\in\mathbb{N}_1^p\}$, a barrier function is defined as
	\begin{equation*}\label{Eqn:barrier_in}
	\bar {\mathcal{B}}(z)=\left\{\begin{array}{l}
	-\sum_{i=1}^p{\rm{log}}\big(1- g_i(z)\big)\ \ z\in \rm {Int}(\mathcal{Z})\\
	[0.2cm]+\infty\ \ \rm{otherwise}.
	\end{array}\right. \qquad
	\end{equation*}\hfill $\blacktriangleleft$
\end{definition}
 
 We introduce the relaxed barrier function in the following lemma according to~\cite{wills2004barrier,nocedal2006numerical}.

\begin{lemma}[Relaxed barrier function] \label{lemma:relax}\hfill
  Let $g_i(z)=a_i^{\top}z$ where $a_i\in\mathbb{R}^{\bar n}$, and define a relaxed barrier function as
		\begin{equation}\label{Eqn:relaxed_B}
		{\mathcal{B}}(z)=\left\{\begin{array}{ll}
		\bar {\mathcal{B}}(z)&\bar\sigma\geq \kappa\\
		\gamma_b(z,\bar\sigma)&\bar\sigma <\kappa
		\end{array}\right.
		\end{equation}
		where $\kappa>0$ is a relaxing factor, 
		$\bar\sigma=\min_{i\in\mathbb{N}_{i=1}^p}1-a_i^{\top}z$, the function $\gamma_b(z,\bar\sigma):(-\infty,\kappa)$ is strictly monotone and differentiable, and $\bigtriangledown^2\gamma_b(z,\bar\sigma)\leq\triangledown^2{\mathcal{B}}(z)|_{\bar\sigma= \kappa}$,  then there exists a positive-definite matrix $H_z\geq\frac{1}{2}\triangledown^2{\mathcal{B}}(z)|_{\bar\sigma= \kappa},$ such that ${\mathcal{B}}(z)\leq z^{\top}H_zz\leq {\mathcal{B}}_{\rm max}(z),$ where ${\mathcal{B}}_{\rm max}(z)=\max_{z\in\mathcal{Z}}z^{\top}H_zz$. \hfill $\blacklozenge$
\end{lemma}
\textbf{Proof}: In view of Definition~\ref{defi:polytopic}, for $\bar\sigma\geq \kappa$, one has  $\bigtriangledown^2\gamma_b(z,\bar\sigma)\leq\bigtriangledown^2 \mathcal{B}(z)\leq 2H_z$. As $\mathcal{B}(0)=0$, it holds that  $\mathcal{B}(z)\leq z^{\top}H_zz$ is verified.
\hfill $\square$
{\color{black}
	\begin{definition} [Feasible control policy]\label{def:2}
		The control policy	$\bm {\hat u}_{k:k+N-1}$ is feasible in the prediction interval $[k,k+N-1]$ if $\hat u(\tau|k)\in\hat{\mathcal{U}}$,  $\hat z(\tau|k)\in\mathcal{Z}$, and $\hat z(k+N|k)\in\mathcal{Z}_f$, $\forall\tau\in\mathbb{N}_0^{N-1}$.
\end{definition}}
\section{Design of robust MPC based on RL}
 
 In this section, we first give the control structure of the proposed robust MPC based on RL, i.e., r-LPC. Then we present by order the finite-horizon RL algorithm and the actor-critic implementation, for learning an explicit nominal control policy of r-LPC  in each prediction interval. Finally, the pseudocode of r-LPC is given at the end of this section. 
 
\subsection{Control structure of r-LPC}
\begin{figure}[h]
	\centering
	\includegraphics[width=0.4\textwidth]{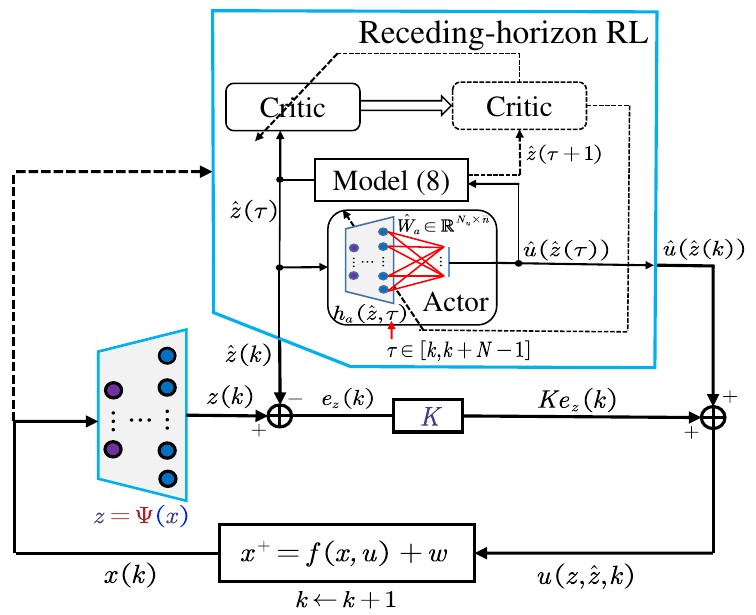}
	\caption{The control structure of the proposed r-LPC algorithm.}
	\label{fig:control_diagram-2}\vspace{0mm}
\end{figure}

\begin{figure}[h]
	\centering
	\includegraphics[width=0.35\textwidth]{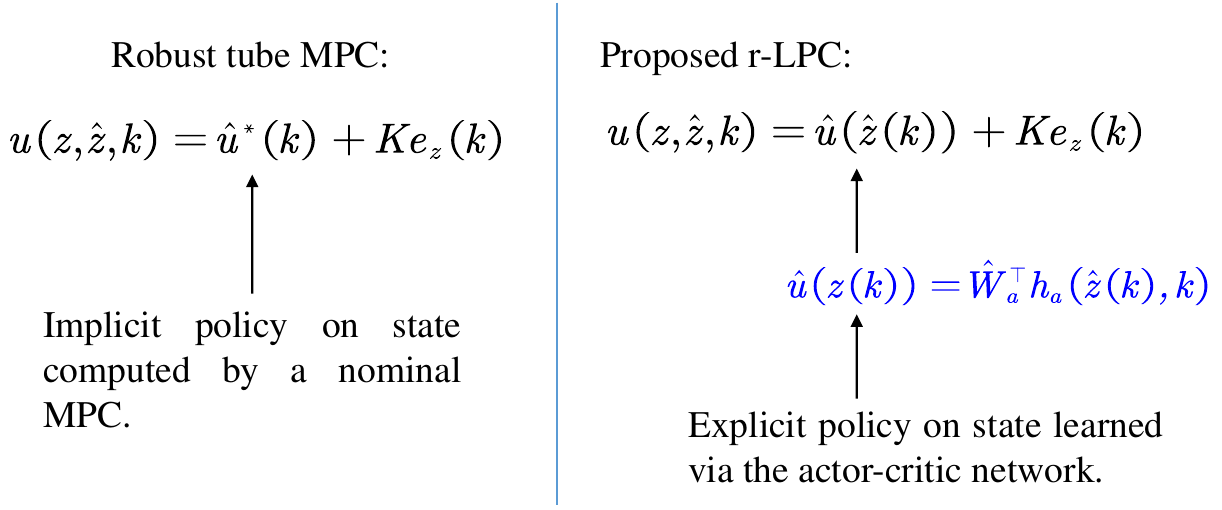}
	\caption{The difference between r-LPC and classic tube MPC. 	In classic robust tube MPC (on the left-hand side), the controller is composed of a state feedback term and an implicit control policy $\hat u^{\ast}$ via numerically solving an online nominal MPC. Whilst, in the proposed r-LPC approach, the term $\hat u(\hat z)$, is an explicit policy on the state, learned via an actor-critic network. 
	}
	\label{fig:control_diagram}\vspace{0mm}
\end{figure}

In the proposed r-LPC approach, the control policy at any time $k$ is of type
\begin{equation}\label{Eqn:real_u_proposed}
u(z,\hat z,k)=\hat u(\hat z(k))+Ke_{\scriptscriptstyle{z}}(k)
\end{equation}
where $\hat u(\hat z)$ is an explicit network-based control policy on the state $\hat z$ learned by a {\color{black}receding-horizon RL} algorithm using an actor-critic structure, while the second term is an off-line nonlinear state-feedback policy, see Fig.~\ref{fig:control_diagram-2}.
Peculiarly, in each prediction interval $[k,k+N-1]$ of r-LPC, the actor and critic adopted in the finite-horizon RL (deferred in Section~\ref{sec:actor-critic}), are designed with neural networks for approximating the near-optimal control policy and value function. The learned actor-network results  in a nominal control policy, i.e., ${\hat u}(\hat z(k+i|k))$, $i=0,\cdots,N-1$.
Then the resulting control to be applied at time $k$ is
\begin{equation}\label{Eqn:con-applied-learned}
u(z,\hat z,k)=\hat u(\hat z(k|k))+Ke_{\scriptscriptstyle{z}}(k).
\end{equation}
At the next time $k+1$, the control policy $\hat u(\hat z)$ can be directly applied or improved via repeatedly solving the finite-horizon learning problem. 
Also, the difference of the proposed r-LPC with classic robust tube-based MPC is highlighted in Fig.~\ref{fig:control_diagram}.
%
In the following two subsections, we focus on presenting the finite-horizon constrained RL with an actor-critic structure for learning the nominal control policy, i.e., $\hat u(\hat z)$ in the prediction interval.  First, the constrained Hamilton-Jacobi-Bellman (HJB) equation is formulated and a finite-horizon iterative RL is given to solve the HJB equation. Then, the actor-critic structure is used to implement the finite-horizon RL, which results in a network-based nominal control policy.
\subsection{Constrained finite-horizon RL in the prediction interval}
Note that, it is nontrivial to deal with state constraints in the learning process. Hence, {\color{black}we first transform the hard state constraint $\hat z\in\mathcal{Z}$ and control constraint $\hat u\in\hat{\mathcal{U}}$  into soft ones with barrier functions defined in Section~\ref{sec:nece-defin}. Specifically, in line with~\cite{wills2004barrier,feller2017relaxed}, the value function in~\eqref{Eqn:optimizL} is reconstructed using barrier functions on the states and controls, i.e., 
\begin{equation}\label{Eqn:V_bz}
\begin{array}{ll}
\bar V\big(\hat{z}(k)\big)= 
\sum_{i=0}^{N-1}
r(\hat{z}(k+i),\hat u(k+i))+\bar V_f\big(\hat{z}(k+N)\big)
\end{array}
\end{equation}
where $r(\hat{z}(\tau),\hat u(\tau))=\|\hat{z}(\tau)\|_{\bar Q}^2+
\|\hat u(\tau)\|_{R}^2+\mu \mathcal{B}(\hat {z}(\tau))+\mu \mathcal{B}(\hat u(\tau))$, $\bar V_f\big(\hat{z}(k+N)\big)=V_f(\hat {z}(k+N))+\mu \mathcal{B}_f(\hat{z}(k+N))$,
 $\mu>0$ is a weighting scalar which determines the influence of barrier function values on $\bar V\big(\hat{z}(k)\big)$,} barrier functions $\mathcal{B}(\hat{z})$, $\mathcal{B}(\hat u)$, and $\mathcal{B}_f(\hat {z})$ are  designed according to Definition~\ref{defi:polytopic} and Lemma~\ref{lemma:relax}. In the case that  ${\mathcal{Z}}_{f}$ is of an ellipsoidal form, i.e., ${\mathcal{Z}}_{f}=\{z\in\mathbb{R}^{\bar n}|g(z)\leq 1\}$  and $g(z)$ is a quadratic function,  we set $\mathcal{B}_f(\hat {z})=\bar{\mathcal{B}}(\hat {z})$. 
 
In view of~\eqref{Eqn:V_bz}, the Lyapunov equation in~\eqref{Eqn:LYA} is modified as
\begin{equation}\label{Eqn:qn_ph}
F^{\top}PF-P=-\bar Q-K^{\top}RK-\mu H,
\end{equation}
where $H=H_{\hat{z}}+K^{\top}H_{\hat u}K$, $H_{\hat{z}}$ and $H_{\hat u}$ are computed  according to Lemma~\ref{lemma:relax} for constraints $\mathcal{Z}$ and $\hat {\mathcal{U}}$ respectively.


{\color{black}With the reconstructed barrier function-based cost function~\eqref{Eqn:V_bz}, the original constrained optimization problem~\eqref{Eqn:optimizL} is transformed as an equality constrained one which can be solved via the RL framework.}
To this end, at any time $k$, {\color{black}letting the remaining cost function at the prediction time $\tau\in[k,k+N-1]$ be $\bar V\big(\hat {z}(\tau)\big)=\sum_{i=0}^{N-\tau-1}
r(\hat{z}(\tau+i),\hat u(\tau+i))+\bar V_f\big(\hat{z}(k+N)\big)$, one can write 
\begin{equation}\label{Eqn:V_bzk}
\begin{array}{ll} \bar V\big(\hat {z}(\tau)\big)=r(\hat z(\tau),\hat u(\tau))+ \bar V\big(\hat{z}(\tau+1)\big)
\end{array}
\end{equation}
where  $\bar V\big(\hat{z}(k+N)\big)=\bar V_f\big(\hat{z}(k+N)\big)$.}

Let $\bar V^{\ast}\big(\hat{z}(\tau)\big)$ be the optimal value function at {\color{black} $\tau\in[k,k+N-1]$}, then the discrete-time HJB equation is given as
\begin{equation}\label{Eqn:u_optimal}
\bar V^{*}(\hat{z}(\tau))= \min_{\hat u(\tau)}  r(\hat{z}(\tau),\hat u(\tau))+ \bar V^{\ast}\big(\hat{z}(\tau+1)\big)\end{equation}
leading to the optimal control policy
\begin{equation}\label{Eqn:u_optimal-argmin}
\hat u^{\ast}(\hat z(\tau))={\rm argmin}_{\hat u(\tau)}   r(\hat{z}(\tau),\hat u(\tau))+ \bar V^{\ast}\big(\hat{z}(\tau+1)\big)\end{equation}
{\color{black}Let $\lambda(\hat{z}(\tau))=\partial \bar V(\hat{z}(\tau))/\partial \hat z(\tau)$ be the costate and denote $\lambda_f(\hat{z}(k+N))=\lambda(\hat{z}(k+N))$ as the terminal costate.}
Substituting $\hat u$ with  $\hat u^{\ast}$ into~\eqref{Eqn:u_optimal} and in view of the optimality condition $\partial \bar V^{\ast}(\hat{z}(\tau))/\partial \hat u^{\ast}(\hat z(\tau))=0$, one has 
\begin{equation}\label{Eqn:hjb}
\mu\frac{\partial \mathcal{B}(\hat u^{\ast}(\hat z(\tau)))}{\partial \hat u^{\ast}(\hat z(\tau))}+2R\hat u^{\ast}(\hat z(\tau))+B^\top \lambda^{\ast}(\hat{z}(\tau+1))=0.
\end{equation}
From~\eqref{Eqn:u_optimal}, one has
\begin{equation}\label{Eqn:lambda}
\begin{array}{ll}
\lambda^{\ast}(\hat{z}(\tau))
=\mu\frac{\partial \mathcal{B}(\hat{z}(\tau))}{\partial \hat{z}(\tau)}+2\bar Q\hat{z}(\tau)+A^\top \lambda^{\ast}(\hat{z}(\tau+1)).
\end{array}
\end{equation}

The computation of $\hat u^{\ast}$ with~\eqref{Eqn:u_optimal} and~\eqref{Eqn:u_optimal-argmin} results in a nonlinear problem since $\mathcal{B}(\cdot)$ is nonlinear.  To solve this problem, the following finite-horizon iterative RL (i.e., value iteration) is introduced.  Given an initial choice $\lambda^{0}(\hat{z}(\tau))=0$, the following two main steps are repetitively performed for all $\tau=k,\cdots, k+N-1$.
\begin{enumerate}[Step 1.]
	\item Policy improvement: provided $\lambda^i(\hat{z}(\tau+1))$,
	\begin{equation}\label{Eqn:policy-im}
	\begin{array}{ll}
\hat u^{i+1}(\hat z(\tau))=&\hspace{0mm}
{\rm arg}_{\hat u(\tau)}\{\mu\frac{\partial \mathcal{B}(\hat u(\tau))}{\partial \hat u(\tau)}+2R\hat u(\tau)+\vspace{1mm}\\
&\hspace{0mm} B^\top \lambda^i(\hat{z}(\tau+1))=0\}
	\end{array}
	\end{equation}
	\item Value function update: given $\hat u^i(\hat z(\tau))$, 
	\begin{equation}\label{Eqn:value-up}
\lambda^{i+1}(\hat{z}(\tau))
=\mu\frac{\partial \mathcal{B}(\hat{z}(\tau))}{\partial \hat{z}{(\tau)}}+2\bar Q\hat{z}(\tau)+A^\top \lambda^i(\hat{z}(\tau+1))
	\end{equation}
\end{enumerate}
\begin{remark}
	The convergence proof of $\hat u^{i}(\tau)$ and $\lambda^{i}(\tau)$ {\color{black}  $\forall \tau \in[k,k+N-1]$} to the optimal values under value iteration steps~\eqref{Eqn:policy-im} and~\eqref{Eqn:value-up} can be easily verified as $i\rightarrow +\infty$, following the results in~\cite{xu2018learning} and~\cite{heydari2012finite}. \hfill \hfill $\blacktriangle$
\end{remark}
 
\subsection{Learning nominal control policy with  actor-critic structure}\label{sec:actor-critic}
In the prediction time $\tau\in[k,k+N-1]$, instead of directly solving steps~\eqref{Eqn:policy-im} and~\eqref{Eqn:value-up}, a regularized actor-critic structure is used where the critic network is to approximate the optimal costate $\lambda^{\ast}(\hat z(\tau))$ and the actor-network is in charge of learning the optimal control policy $\hat u^{\ast}(\hat z(\tau))$. The resulting nominal control policy is nonlinear, network-based, and explicit, which is different from that in classic tube MPC.

We first design the critic network. To this end, let $\lambda^{\ast}(\hat{z}(\tau))$ be represented as $$\lambda^{\ast}(\hat{z}(\tau))=W_{c}^{\top}h_c(\hat{z}(\tau),\tau)+\bar{\epsilon}_c(\tau),$$
where $W_c\in\mathbb{R}^{N_c\times \bar n}$  is the weighting matrix, $h_c\in\mathbb{R}^{N_c}$ is  a vector of basis functions, $\bar{\epsilon}_c(\tau)$ is the network residual.
Let the adopted critic network be defined as
\begin{equation}\label{eqn:critic}
\hat\lambda(\hat{z}(\tau))=\hat W_c^{\top}h_c(\hat{z}(\tau),\tau),
\end{equation}
for all $\tau\in[k,k+N]$, {\color{black} where $\hat\lambda(\hat{z}(k+N))$ is the approximated terminal costate, i.e., $\hat\lambda(\hat{z}(k+N))=\hat\lambda_f(\hat{z}(k+N))$,} $\hat W_c$ is the approximation of $W_c$ via minimizing the distance of the optimal costate $\lambda^{\ast}$ and $\hat\lambda$. Note however $\lambda^{\ast}$ is not available, in line with~\cite{heydari2012finite} and~\eqref{Eqn:value-up}, we define a target to be steered by $\hat \lambda$, i.e.,
\begin{equation}\label{Eqn:lam_d}
\begin{array}{l}
\lambda_d(\hat {z}(\tau))=\\
\left\{\begin{array}{l}\mu \frac{\partial \mathcal{B}(\hat {z}(\tau))}{\partial \hat {z}(\tau)}+
2\bar Q\hat {z}(\tau)+
A^\top \hat\lambda(\hat {z}(\tau+1)), \ \tau\in[k,k+N)\vspace{1mm}\\
\mu \frac{\partial \mathcal{B}_f(\hat {z}(\tau))}{\partial \hat {z}(\tau)}+
2P\hat{z}(\tau):=\lambda_{d,f}(\hat{z}(\tau)),\ \tau=k+N
\end{array}\right.
\end{array}
\end{equation}
where $\hat{z}(k+N)$ can be randomly chosen from set ${\mathcal{Z}}_{f}$. 

\begin{remark}
	In the learning process, with the goal of steering $\lambda_d$,
 $\hat\lambda$ is recursively updated along with $\hat W_c$ based on the gradient descend mechanism, which however might lead to the state constraint nonsatisfaction since $\hat W_c$ is related to $\hat \lambda$ (rather than $\hat z$). To improve the learning efficiency, one can instead enforce the restriction on $\hat \lambda$. To proceed, let at time $\tau=k+j$, $\bar {\mathcal{Z}}^j$ be the feasible set of problem~\eqref{Eqn:optimizL}, which can be computed according to~\cite{scibilia2011feasible} since Assumption~(A1) holds. With $\bar {\mathcal{Z}}^j$, one can compute an estimate of the set where $\hat \lambda(\tau)$ lies in, i.e., $\Lambda^j=2\bar Q\bar {\mathcal{Z}}^j+A^{\top}\Lambda^{j+1}$, starting from $\Lambda^{N}=2P\Lambda^N$, since $\partial \mathcal{B}(\hat {z})/\partial \hat {z}$ is close to 0 in view of~\eqref{Eqn:lam_d}. \hfill $\blacktriangle$
\end{remark}

   Once $\Lambda^j$, $\forall j\in\mathbb{N}_1^N$ are available, {\color{black}letting $\epsilon_{c}(\tau)=\lambda_d(\hat{z}(\tau))-\hat\lambda(\hat{z}(\tau))$ $\forall\tau\in[k,k+N-1]$ and $\epsilon_{c,N}=\lambda_d(\hat{z}(k+N))-\hat\lambda(\hat{z}(k+N))$,} it is possible to define a regularized optimization cost of the critic network as
\begin{equation}\label{Eqn:ec}
\delta_c(\tau)=\|(\epsilon_{c}(\tau),\epsilon_{c,N})\|^2+\bar\mu (\mathcal{B}(\hat\lambda(\hat {z}(\tau)))+\mathcal{B}_f(\hat\lambda(\hat z(k+N)))),
\end{equation}
 where $\mathcal{B}(\cdot)$ and $\mathcal{B}_f(\cdot)$ are regularization terms  on $\hat W_c$, defined as barrier functions of constraints  $\hat\lambda(\hat{z}(k+j))\in\Lambda^j$ and $\hat\lambda(\hat{z}(k+N))\in\Lambda^N$.
At any time instant $\tau$, the  weight $\hat W_{c}$ is updated via minimizing~\eqref{Eqn:ec}  according to the following rule

\begin{equation}\label{Eqn:wc}
\hat W_{c}(\tau+1)={\hat W_{c}(\tau)} - {\gamma_c}\frac{\partial\delta_c(\tau)}{\partial \hat W_{c}(\tau)},
\end{equation}
where $\gamma_c$ is the learning rate. {\color{black}Note that, the term $\partial\delta_c(\tau)/\partial \hat W_{c}(\tau)=\partial\delta_c(\tau)/\partial \hat \lambda(\tau)\cdot \partial\hat\lambda(\tau)/\partial \hat W_{c}(\tau)$ can be computed according to the chain rule since $\epsilon_{c}(\tau)$ and $\epsilon_{c,N}$ are related to $\hat\lambda$, and $\hat\lambda$ is a function on $\hat W_c$ (see~\eqref{eqn:critic}).}

Likewise, $\hat u^{\ast}(\hat z(\tau))$ can be represented using a network: $$\hat u^{\ast}(\hat z(\tau))=W_{a}^{\top}h_a(\hat{z}(\tau),\tau)+\bar{\epsilon}_a(\tau),$$
where $W_{a}\in\mathbb{R}^{N_u\times m}$  is the weighting matrix, $h_a\in\mathbb{R}^{N_u}$ is a vector of basis functions, $\bar{\epsilon}_a(\tau)$ is the network residual.
To define the actor network, in view of ~\eqref{Eqn:hjb}, we define a desired control action $ \hat u_d(\hat z(\tau))$, $\tau\in[k,k+N-1]$ satisfying
\begin{equation}\label{Eqn:act-d}
\begin{array}{ll}
G(\hat u_d(\hat z(\tau)))&:=\mu\frac{\partial \mathcal{B}( \hat u_d(\hat z(\tau)))}{\partial \hat u_d(\hat z(\tau))}+2R\hat u_d(\hat{z}(\tau))\\
&\ =-B^\top \hat\lambda(\hat{z}(\tau+1)).
\end{array}
\end{equation}
The approximation of $\hat u_d(\hat z(\tau))$ is then defined as
\begin{equation}\label{Eqn:actor}
\hat u(\hat z(\tau))=\hat W_{a}^{\top}h_a(\hat{z}(\tau),\tau), \tau\in[k,k+N-1],
\end{equation}
where $\hat W_{a}$  is an approximation of $W_{a}$.

Note that, the left-side of~\eqref{Eqn:act-d} is composed of a linear mapping and a nonlinear function on $\hat u_d$.
Letting
$\epsilon_{a}(\tau)=G(\hat u_d(\hat{z}(\tau)))-G(\hat u(\hat{z}(\tau)))$, {\color{black}$\|\epsilon_{a}(\tau)\|^2$ is to be minimized in the learning process, which is equivalent to driving $
\mu \partial \mathcal{B}\left( \hat{u}\left( \hat{z}\left( \tau \right) \right) \right)/\partial \hat{u}\left( \hat{z}\left( \tau \right) \right)+2R\hat{u}\left( \hat{z}\left( \tau \right) \right) +B^{\top}\hat{\lambda}\left( \hat{z}\left( \tau +1 \right) \right) 
\rightarrow 0$ (see~\eqref{Eqn:hjb}) in the Euclidean norm sense.} To also enforce a regularization on $\hat{u}$, we choose to penalize $\epsilon_{a}$  
 via a regularized loss function as follows:
\begin{equation}\label{Eqn:loss_a}
\delta_{a}(\tau)=\|\epsilon_{a}(\tau)\|^2+\bar\mu \mathcal{B}(\hat u(\hat z(\tau))),\\
\end{equation}
where the second term imposes a regularization on $\hat W_a$ with barrier functions.
At any time instant $\tau\in[k,k+N-1]$, $\hat W_{a}$ is updated via minimizing $\delta_{a}(\tau)$ under the following rule:
\begin{equation}\label{Eqn:wa}
\hat W_{a}(\tau+1)={\hat W_{a}(\tau)} - {\gamma_a}\frac{\partial\delta_{a}(\tau)}{\partial \hat W_a(\tau)},
\end{equation}
where  $\gamma_a$ is the learning rate. {\color{black}Likewise, the term $\partial\delta_a(\tau)/\partial \hat W_{a}(\tau)={\partial\delta_{a}(\tau)}/{\partial \hat u(\tau)}\cdot{\partial \hat u(\tau)}/{\partial \hat W_a(\tau)}$ can be computed according to the chain rule since $\epsilon_{a}$ is related to $\hat u$ and $\hat u$ is a function on $\hat W_a$ (see~\eqref{Eqn:actor}). }

  \begin{algorithm}
	\caption{Pseudocode of r-LPC.}\label{Atm:2}
		\textbf{Off-line designs:}\vspace{0mm}\\
	{\color{black}	\begin{algorithmic}[1]
	\State  choose $\Psi(x)$ such that Assumptions (A2) and (A3) are fulfilled;
	\State  calculate $A$, $B$, $C$, and $D$ with~\eqref{Eqn:appro_K} (see Appendix~\ref{SEC:COM-SETS}), and check that Assumption (A1) is satisfied;
	\State  compute ${\mathcal{D}}$ and $\mathcal{V}$ according to~\cite{XLZ2021};
	\State  calculate $P$ and $K$ with~\eqref{Eqn:qn_ph} and compute $\mathcal{O}_z$, $\mathcal{O}_x$ with $K$;
	\State  compute  $\mathcal{Z}$, $\hat {\mathcal{U}}$, and $\mathcal{Z}_f$ with $K$.\vspace{0mm}
	\end{algorithmic}}
	\dotfill \\
\textbf{Online procedures:}
	\begin{algorithmic}[1] 
		\Require Set the iteration threshold $\bar i$, initialize  $\gamma_a$, $\gamma_c$, select $\hat W_{a}(\tau)$ and $\hat W_{c}(\tau)$ such that $\delta_{a}<\mathcal{B}_{\rm max}(\hat u)$, and $\delta_c<\mathcal{B}_{\rm max}(\hat\lambda)$, $i=0$.\\
		{ At any time steps $k=1, 2, \cdots$ \textbf{do}}\\
\dotfill \emph{\%\%learning process in the prediction interval}\dotfill
		\Repeat \, 			\emph{\%\%iteration loop in the prediction interval}
		\For{$\tau=k, \cdots, k+N-1$}\, \emph{\%\%finite-horizon forward-in-time learning} 
		\State compute $\hat u(\hat z(\tau))$, $\hat {z}(\tau+1)$ with~\eqref{Eqn:actor},~\eqref{Eqn:unpert};
		\State compute $\hat\lambda(\hat{z}(\tau+1))$, $\lambda_d(\hat {z}(\tau))$ with~\eqref{eqn:critic},~\eqref{Eqn:lam_d}
		\State 	compute $G(\hat u_d(\hat{z}(\tau)))$ with~\eqref{Eqn:act-d};
		{\color{black}\State  generate  $\hat\lambda(\hat{z}(k+N))$, $\lambda_d(\hat {z}(k+N))$ with~\eqref{eqn:critic},~\eqref{Eqn:lam_d}  using randomly selected terminal state $\hat{z}(k+N)\in\mathcal{Z}_f$;
		\State	compute $\epsilon_{c}(\tau),\epsilon_{c,N}$ and update $\hat W_{c}(\tau+1)$ with~\eqref{Eqn:ec}, \eqref{Eqn:wc};
		\State	compute  $\epsilon_{a}(\tau)$ and update  $\hat W_{a}(\tau+1)$ with~\eqref{Eqn:loss_a}, \eqref{Eqn:wa};}
		{\color{black}\If{$\hat {z}(\tau)\notin\mathcal{Z}$ $\vee$ $\hat {u}(\tau)\notin\hat{\mathcal{U}}$ $\vee$ $\hat {z}(k+N)\notin\mathcal{Z}_f$} 
		\State Repeat steps 3 to 16 using re-initialized weight~\eqref{Eqn:safe-weight} and break;
		\EndIf}
		\EndFor
		\State $i\leftarrow i+1$
		\Until{$i= \bar i$}\\
			\dotfill \emph{\%\%apply the learned control to the system}\dotfill 
		{\color{black}\If {$\bar V(k|k)\leq \bar V(k|k-1)$}
		\State generate $\hat u(\hat z(k|k))$ with~\eqref{Eqn:actor};
		\Else
		\State compute $\hat u(\hat z(k|k))$ with~\eqref{Eqn:safe-policy};
		\EndIf}
		\State apply $u(z,\hat z,k)=\hat u(\hat z(k|k))+Ke_z(k)$ to system~\eqref{Eqn:non-model}
		\State update $\hat{z}(k+1)$, ${z}(k+1)$, $\hat x(k+1)$, and $x(k+1)$;
		\State $k\leftarrow k+1$;
	\end{algorithmic}
\end{algorithm}
In summary, the pseudocode of the proposed r-LPC approach is given in Algorithm~\ref{Atm:2}.
\section{Main theoretical results}\label{sec:con-actor-critic}
In this section, we first prove the convergence of the barrier-function regularized actor-critic learning algorithm under~\eqref{Eqn:wc} and~\eqref{Eqn:wa} in each prediction interval, which has not been proven by existing works, e.g.,~\cite{Xu-finitehorizonadp,zhao2014neural,xu2018learning,dong2018functional}. Then we give a necessary condition to guarantee the recursive feasibility and robustness of r-LPC, which is the main focus of our work. Moreover, we show that the closed-loop asymptotic stability can be ensured under no external disturbance. 

We first introduce the following assumptions.
	\begin{enumerate}
		\item[{(A4)}]\label{Eqn:bound-network}  $\|W_{\star}\|\leq W_{\star,m}$, $\|h_{\star}\|\leq h_{\star,m}$, $\|\bar \epsilon_{\star}\|\leq \bar \epsilon_{\star,m}$, where $\star=a,c$ in turns. \hfill $\blacktriangleleft$
	\end{enumerate}

\begin{theorem} [Convergence of actor-critic]\label{the:ac-conver} Under Assumption~(A4) and the learning rules~\eqref{Eqn:wc},~\eqref{Eqn:wa}, if learning rates $\gamma_a$ and $\gamma_c$  
	are such that	$\bar G_{c1}, \bar G_{a1}>0$ (whose definitions are deferred in Appendix \ref{sec:proof_2}),  then there exist $\eta_{c},\eta_{a}\geq0$ such that
	\begin{equation}\label{Eqn:ac-uub}
	\begin{array}{ll}
	\|\hat\lambda(\hat{z}(\tau+1))-\lambda^{\ast}(\hat{z}(\tau+1))\|\leq \eta_{c}\vspace{1mm}\\
	\|\hat u(\hat z(\tau))-\hat u^{\ast}(\hat z(\tau))\|\leq \eta_{a},
	\end{array}
	\end{equation}
	$\tau \in[k,k+N-1]$, as $N\rightarrow+\infty$.
	 Also, if 
	 $\bar\epsilon_{a},\bar \epsilon_{c},\bar \mu\rightarrow 0$, and $A$ is Schur stable, then
		\begin{equation}\label{Eqn:ac-con}
	\begin{array}{ll}
\hat W_c(\tau)\rightarrow W_c, \quad \hat W_a(\tau)\rightarrow W_a
	\end{array}
\end{equation} 
and
\begin{equation}
\begin{array}{ll}
\hat\lambda(\hat{z}(\tau+1))\rightarrow\lambda^{\ast}(\hat{z}(\tau+1))\vspace{1mm}\\
\hat u(\hat z(\tau))\rightarrow\hat u^{\ast}(\hat z(\tau)),
\end{array}
\end{equation}
$\tau \in[k,k+N-1]$, as  $N\rightarrow+\infty$. \hfill$\blacksquare$
\end{theorem}
Proof: please refer to Appendix~\ref{sec:proof_2}. \hfill$\square$
\begin{remark}
	Theorem~\ref{the:ac-conver} implies that  a sufficient large choice of prediction horizon $N$ is required to guarantee~\eqref{Eqn:ac-uub}. To achieve this, an outer iterative loop with a length of $\bar i$ (see Algorithm~\ref{Atm:2}) can be adopted since the weighting matrices $\hat W_c$ and $\hat W_a$ are time-independent. This allows using a smaller choice of $N$ via increasing $\bar i$, such that $\bar iN$ is sufficiently large. 
	
	\hfill $\blacktriangle$ 
\end{remark}


To state the following theorem in a compact form, letting $\hat u(\hat z(\tau|k))$ be a control policy at any time $k$, we define a backup policy at time $k+1$ as
\begin{equation}\label{Eqn:safe-policy}
\begin{array}{ll}
\bm {\hat u}^s(\hat z(k+1))=\\
\hat u(\hat z(k+1|k)),\cdots, \hat u(\hat z(k+N-1|k)), K\hat z(k+N|k)
\end{array}
\end{equation}
 Also,  we let ${\bm h}_a=[{h}_a(\hat z^s(k+1)),\,\cdots,\, {h}_a(\hat z^s(k+N))]$, where $\hat z^s$ are computed with control policy $\bm {\hat u}^s$.
\begin{theorem}[Recursive feasibility]\label{the:2} Under Assumption (A1)-(A4), if the learned control policy $\hat u(\hat z(\tau|k))$ is feasible (see Definition~\ref{def:2}) at time $k$, then $\bm {\hat u}^s(\hat z(k+1))$
	 is feasible  at time $k+1$. 
	 Also, if ${\bm h}_a$ satisfies the persistent excitation condition, i.e.,
	 \begin{equation}\label{Eqn:persistent}
	 \rho_1I\leq {\bm h}_a{\bm h}_a^{\top}\leq \rho_2 I
	 \end{equation} where $\rho_1,\rho_2>0$, then the weight $\hat W_a$ satisfying
	 	\begin{equation}\label{Eqn:safe-weight}
	 \hat W_a^{\top}=\bm {\hat u}^s(\hat z(k+1)){\bm h}_a^{\top}({\bm h}_a{\bm h}_a^{\top})^{\dagger}
	 \end{equation}
	 constructs a feasible control policy in the interval $[k,k+N-1]$.  
	 
	 \hfill$\blacksquare$
\end{theorem}
Proof: please refer to Appendix~\ref{appen:the2}.
\hfill$\square$
\begin{remark}
	Note that $\bm {\hat u}(\hat z)=\bm {\hat u}^s(\hat z)$ can be directly used whenever a learning failure occurs. For the sake of control performance improvement, one can use $\hat W_a$~\eqref{Eqn:safe-weight} as a feasible initialization for improving the control policy.  \hfill $\blacktriangle$
\end{remark}
\begin{remark}
	{\color{black}We highlight that recent work in~\cite{han2021reinforcement} proposed a Lyapunov-based model-free RL approach to control dynamical systems with stochastic safety constraints. The control safety was guaranteed with probability, and the reliability of the safety guarantee was related to the data samples used in the training process. In our case, deterministic constraint satisfaction is achieved by resorting to the receding horizon principle.} 
	\hfill $\blacktriangle$
\end{remark}

\begin{theorem}[Closed-loop robustness]\label{the:3} Under Assumption (A1)-(A4), let at time $k$
	\begin{equation}\label{Eqn:safe-u}
	\hat u(k)=\left\{\begin{array}{lll}
	{\hat u}(\hat z(k|k))\ \ \ \  {\text{if}\ \bar V(k|k)\leq \bar V(k|k-1)}\\
	{\hat u}(\hat z(k|k-1))\ \ \ \ \ \ \ {\rm otherwise},
	\end{array}\right.
	\end{equation} 
	then the closed-loop asymptotic convergence of $\hat u$, $\hat z$, $\hat x$ to the origin and of $u$, $z$, $x$ to the robust  tubes are verified, i.e., $\hat u(k),\hat z(k),\hat x(k)\rightarrow 0,$
	and
	$ u(k)\rightarrow K{\mathcal{O}}_z,\  z(k)\rightarrow {\mathcal{O}}_z,\ x(k)\rightarrow {\mathcal{O}}_x\ \text{as} \ k\rightarrow+\infty.$ \hfill$\blacksquare$
\end{theorem}
Proof: please refer to Appendix~\ref{appen:the3}. \hfill$\square$

In the following, we prove the asymptotic stability under $w=0$. To proceed, in view of~\cite{XLZ2021}, $\Psi^{-1}$ is Lipschitz continuous under Assumption~(A3), then one can derive 	\begin{equation}\label{Eqn:lipsch-1}
\begin{array}{ll}
\|\bar w(z,u,0)\|\leq L_z \|z\|+L_u \|u\|,
\end{array}
\end{equation}
for all $z=\Psi(x)$ satisfying $x\in\mathcal{X}$,  $u\in\mathcal{U}$, where $L_z$ and $L_u$ are Lipschitz constants.  From~\eqref{Eqn:lipsch-1}, one has $\|\bar w\|\leq L_z \|z\|+L_u \|z\|_{K^{\top}K}$ in view of~\eqref{Eqn:con-applied-learned} and of  $\hat u,\hat z\rightarrow 0$ asymptotically.
\begin{theorem}[Asymptotic stability]\label{the:rmpc-convergence} If $w=0$ and there exists a scalar $\gamma>0$ such that \begin{equation}\label{Eqn:iss-con}
\begin{bmatrix}
F^{\top}PF-P+\gamma\bar L&F^{\top}P\\PF&-\gamma I
\end{bmatrix}\prec0,
	\end{equation}
		where $\bar L=(L_zI+L_uK^{\top}K )$, 
	the  closed-loop system~\eqref{Eqn:linear_p-residual} and~\eqref{Eqn:non-model} with~\eqref{Eqn:real_u} converge to the origin asymptotically, i.e.,
	$x(k)\rightarrow 0,\ \ u(k)\rightarrow 0,\   \text{and}\ z(k)\rightarrow 0$ asymptotically. \hfill$\blacksquare$\\
	Proof: please refer to Appendix~\ref{appen:the:rmpc-convergence}. \hfill$\square$
\end{theorem}
\begin{remark}
	 The interior-point numerical optimization method takes $O(N(\bar n + m)^3
	)$ operations for systems with block-diagonal structure and $O(N^3(\bar n+ m)^3
	)$ operations if the block-diagonal structure is not exploited, see~\cite{2010Fast}. However, the online computational complexity of our method is roughly $O(N(\bar n+m)(\bar n+N_c))$ given $N_c\geq N_a$. Also, considering an off-line training and an online deploying case, the computational load is only due to~\eqref{Eqn:real_u_proposed} and~\eqref{Eqn:actor}. \hfill $\blacktriangle$
\end{remark}

\section{Simulation and experimental results}\label{sec:simulation}

\subsection{Simulation results on Van der Pol oscillator}
Consider a continuous-time Van der Pol oscillator~\cite{korda2018linear}. Its model is given as
\begin{equation}\label{Eqn:ct}
\begin{array}{ll}
\begin{bmatrix}
\dot x_1\\
\dot x_2
\end{bmatrix}=&\begin{bmatrix}
x_1\\
-2x_2+10x_1^2x_2+0.8x_1+u
\end{bmatrix}+w,
\end{array}
\end{equation}
where $x_1$ and $x_2$ are the states, and $u$ is the control, the disturbance $w=0.4{\rm sin} (10\pi t)$. Let $x=(x_1,x_2)$, then the following constraints are restricted, i.e.,
$
-(2.5,2.5)\leq x\leq(2.5,2.5),\ 
-10\leq u\leq 10.$
\begin{figure}[h]\vspace{0mm}
	\centering
	\includegraphics[width=0.35\textwidth]{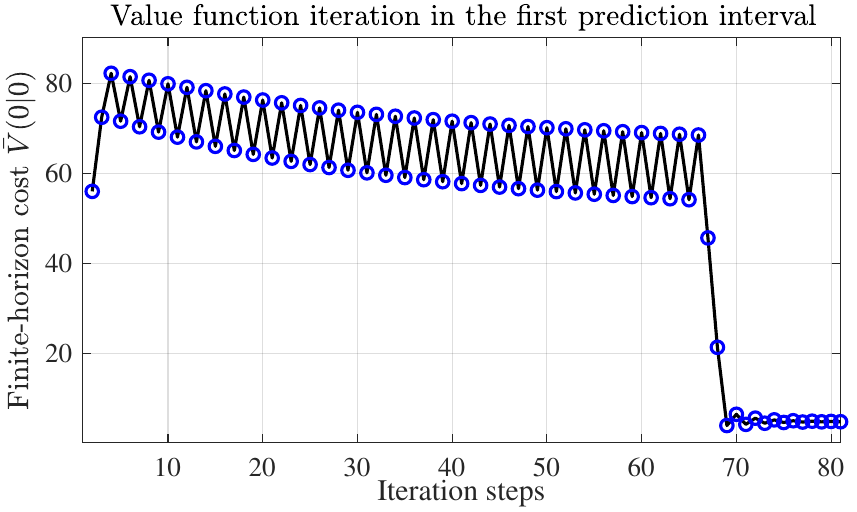}\vspace{0mm}
	\caption{Van der Pol oscillator: variations of the value function $\bar V(0|0)$ in the first prediction interval.}\vspace{0mm}
	\label{fig:cost_com}
\end{figure}
\begin{figure}[h]
	\centering
	\includegraphics[width=0.4\textwidth]{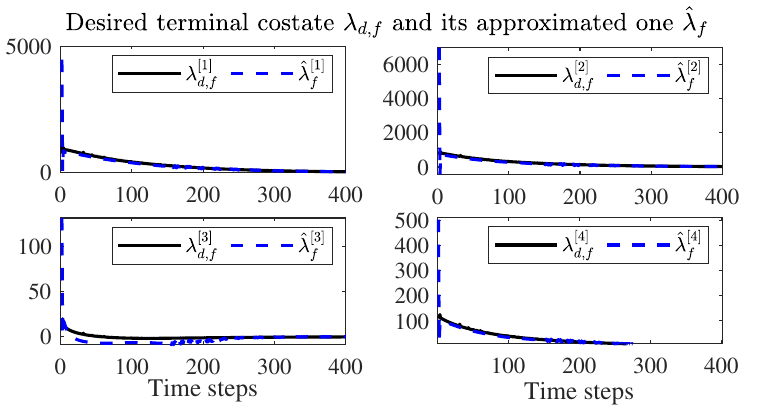}
	\caption{{\color{black}Van der Pol oscillator:  the desired terminal costate  $\lambda_{d,f}$ and the approximated one $\hat{\lambda}_f$, where $\lambda_{d,f}^{[i]}$ ($\hat{\lambda}_f^{[i]}$) is the $i$-th entry of $\lambda_{d,f}$ ($\hat{\lambda}_f$).}}\vspace{0mm}
	\label{fig:vf_van}
\end{figure}
\begin{figure}[h]
	\centering
	\includegraphics[width=0.35\textwidth]{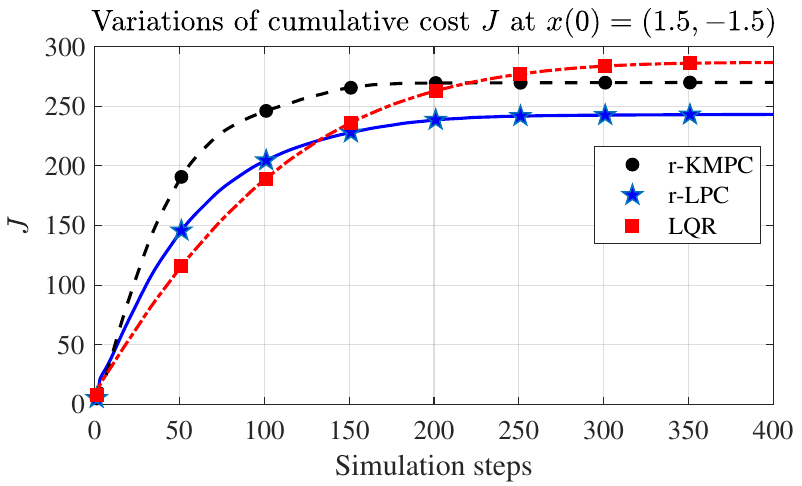}
	\caption{Van der Pol oscillator:  comparison of the cumulative cost $J$ of the three approaches at $x(0)=(1.5,-1.5)$.}\vspace{0mm}
	\label{fig:stage-cost}
\end{figure}
\begin{table*}[h!tb]\vspace{0mm}
	\centering \caption{Van der Pol oscillator: comparisons in terms of cumulative cost $J$.}
	\label{tab:Tab_com0}
	\vskip 0.1cm
	\renewcommand\arraystretch{1.1}
	\scalebox{0.7}{
		\begin{tabular}{cccccccc}
			\toprule
			$x(0)$  & $(0.5,-0.5)$&  $(1,-1)$& $(1,-1.5)$ &$(1.5-1.5)$& $(1.5,-2)$ & $(1.8-1.8)$&$(2,-2)$ \\
			\midrule
			r-LPC (learned online)  &45&70&$\bm {70}$&$\bm {243}$&260&$\bm {528}$&$\bm {848}$\\
			\midrule
			r-LPC (learned off-line) &26&$\bm {66}$&78&280&268&701&889\\
			\midrule
			r-KMPC~\cite{XLZ2021} &$25$ &68&71& 270&$\bm {247}$&599&$861$\\
			\midrule
			LQR &$\bm{22}$ &79&87& 287&275&576&870\\
			\bottomrule\vspace{-8mm}
		\end{tabular}
	}
\end{table*}
 To implement the proposed r-LPC algorithm, model~\eqref{Eqn:ct} was discretized with a sampling period  $T=0.01s$ to obtain the discrete-time counterpart, i.e.,~\eqref{Eqn:non-model}. Then, to derive the Koopman model, we selected a type of $\Psi(x)$ as
$
\Psi(x)=(x,\psi_1(x),\psi_2(x))-(0,\psi_1(0),\psi_2(0))
$ where  $\psi_i(x)=\|x-c_i\|^2\log(\|x-c_i\|),$ $i=1,2$, $\bar n=4$, $c_1=(0.381,
-0.341)$, $c_2=(0.267,-0.889)$ are the kernel centers randomly generated according to a uniform distribution.  The model parameters and sets $\mathcal{V}$, $\mathcal{D}$ are computed according to the steps described in Appendix~\ref{SEC:COM-SETS}. The resulting constraint on the nominal state $\hat x$ was $-[2.38\ 2.05]\leq\hat x \leq [2.38\ 2.05]$.

In the proposed r-LPC, the penalty matrices $Q=I_2$, $R=0.1$. The penalty scalars related to barrier functions were $\mu=\bar \mu=0.001$.  The prediction horizon was $N=10$. The functions $h_a$, $h_c$ were chosen as
$h_a=h_c=(\sigma(\hat z),\nu \tau,\nu \tau^2)$,
where $\sigma(\hat z)= 1 / (1 + {\rm exp}(-W_1\hat z+b_1))$, $\nu=0.001$, where $W_1\in\mathbb{R}^{8\times 4}$, $b_1\in\mathbb{R}^{8}$, are weighting matrices.

\begin{figure}[h]\vspace{0mm}
	\centering
	\includegraphics[width=0.4\textwidth]{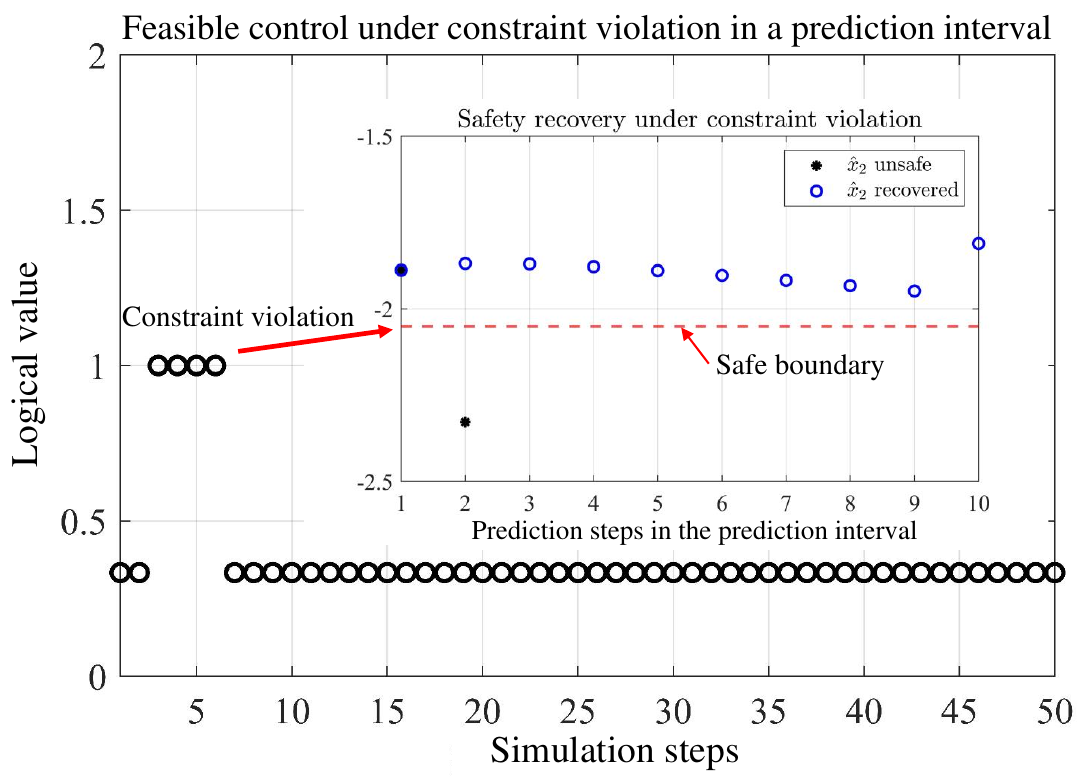}\vspace{0mm}
	\caption{Van der Pol oscillator: feasible control recovery when a constraint violation occurs. The logical value being 1 means a constraint violation is detected when the proposed feasible control policy  is applied to guarantee recursive feasibility of r-LPC.}\vspace{0mm}
	\label{fig:safetytest}
\end{figure}

The proposed r-LPC was first implemented with an initial condition $x(0)=(1.5,-1.5)$. The weighting matrices $\hat W_a$ and $\hat W_c$ were initialized as uniformly distributed random values. The simulations with r-LPC were performed in MATLAB 2019a environment running with Windows 10 operating system. In each prediction interval, multiple episodes were used to learn $\hat W_a$ and $\hat W_c$. The iterative variations of value function $\bar V$ in a prediction interval are displayed in Fig.~\ref{fig:cost_com}, which shows $\bar V$ reduces gradually and converges to the neighbor of a constant value after about 70 iterative steps. 
In the following presented simulation results, the iteration in each prediction interval was terminated if the terminal region was reached and the episode was greater than 20. {\color{black}Also, as shown in Fig.~\ref{fig:vf_van}, the approximated and desired terminal costates converge asymptotically to the origin, and the approximated terminal costate by the critic network is close to the desired one.}

Comparisons with r-KMPC~\cite{XLZ2021} and a linear quadratic regulator (LQR): The comparative studies of r-LPC with  r-KMPC~\cite{XLZ2021} and a linear quadratic regulator (LQR) using a local linearized model were also considered.  For a fair comparison, parameters $Q$ and $R$ in the r-KMPC and LQR were selected the same as r-LPC. 
The variations of cumulative cost $J=\sum_{j=1}^{N_{\rm sim}}\|x(j)\|_Q^2+\|u(j)\|_R^2$, $N_{\rm sim}=400$, of all the three controllers are presented in Fig.~\ref{fig:stage-cost}. The results show that the control performance of r-LPC is slightly better than  r-KMPC and LQR in terms of regulation cost under $x(0)=(1.5,-1.5)$. 
To further verify the effectiveness of r-LPC, we also conducted simulation tests with multiple different initial conditions. The resulting cumulative costs obtained with r-LPC (using online as well as off-line learned policy), r-KMPC, and LQR are listed in Table~\ref{tab:Tab_com0}. The results reveal that r-LPC can perform slightly better in the case that initial conditions were far from the origin. Also, compared with LQR, r-LPC and r-KMPC can deal with constrained control problems. As a piece of evidence to the constraint satisfaction of r-LPC, the result of a safety test is presented in Fig.~\ref{fig:safetytest}, which shows that r-LPC can recover a feasible control policy from constraint violation in the adopted prediction interval.
\vspace{0mm}
\subsection{Simulation and experimental results on an Inverted Pendulum}
\begin{figure}[h]
	\centering
	\includegraphics[width=0.25\textwidth]{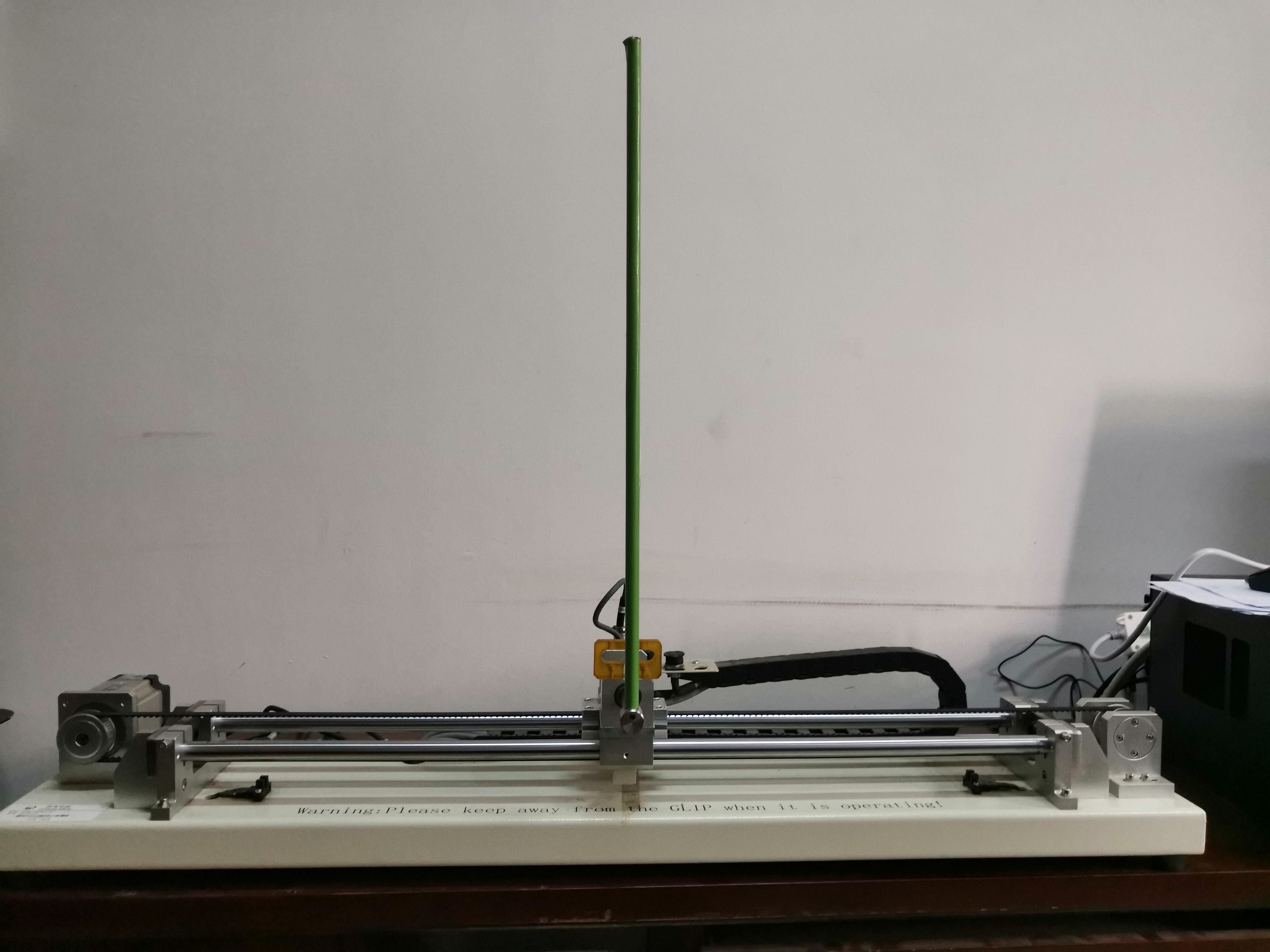}
	\caption{\vspace{0mm}The experimental platform of an inverted pendulum.}
	\label{fig:ip-exp}\vspace{0mm}
\end{figure}

\begin{figure}[h]\vspace{0mm}
	\centering
	\includegraphics[width=0.4\textwidth]{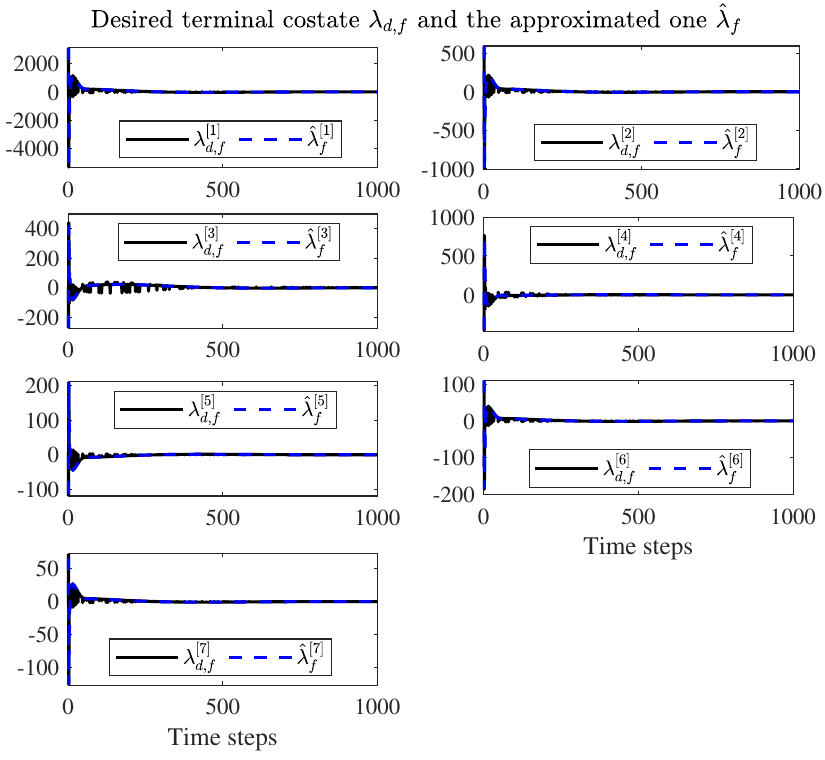}\vspace{0mm}
	\caption{{\color{black}Inverted pendulum: the desired terminal costate  $\lambda_{d,f}$ and the approximated one $\hat{\lambda}_f$, where $\lambda_{d,f}^{[i]}$ ($\hat{\lambda}_f^{[i]}$) is the $i$-th entry of $\lambda_{d,f}$ ($\hat{\lambda}_f$).}}
	\label{fig:vf_ip}
\end{figure}

Consider also the problem of regulating an Inverted Pendulum. The continuous-time nonlinear system model is:
\begin{equation}\label{Eqn:ct-ip}
\begin{bmatrix}
\dot x_1\\
\dot x_2\\
\dot x_3\\
\dot x_4
\end{bmatrix}=\begin{bmatrix}
x_2\\
\frac{3}{2l}(g{\rm sin}x_1+u{\rm cos}x_1)\\
x_4\\
u
\end{bmatrix}+w
\end{equation}
where $x_1$, $x_2$, $x_3$, $x_4$ are the angle displacement and velocity of the rod,  position  and velocity of the car, $l=0.5$ $m$, $g=9.8$ $m/s^2$, $\|w\|_{\infty}\leq 0.5$. Denoting $x=(x_1,x_2,x_3,x_4)$, the state and control are limited as $-(0.25 rad,2 rad/s,1 m,2 m/s)\leq x\leq (0.25 rad,2 rad/s,1 m,2 m/s)$, $|u|\leq 20$$m/s^2$.  For a visual display of the inverted pendulum platform, please see Fig.~\ref{fig:ip-exp}.
To derive the Koopman model, input-output datasets pairs  of~\eqref{Eqn:ct-ip} with $M=2\cdot 10^4$ were collected using a random control policy with a uniform distribution.  Gaussian kernel functions were selected to construct $\Psi$, i.e.,
\begin{equation*}\label{Eqn:phi-choice-ip}
\Psi(x)=(x,\psi_1(x),\psi_2(x),\psi_3(x))-( 0,\psi_1(0),\psi_2(0),\psi_3(0))
\end{equation*} where  $\psi_i(x)=e^{-\|x-c_i\|^2},$ $i=1,2,3$, $\bar n=7$, kernel centers $c_1=(0.59,-0.73,0.14,0.04)$, $c_2=(0.67,0.26,0.2,0.72)$, $c_3=(0.3,0.99,0.58,-0.37)$ are generated randomly.
The parameters of the linear predictor was computed similar to the previous section, see~\cite{XLZ2021}. The penalty matrices $Q$ and $R$ were selected as $Q=I_2$, $R=0.02$.  $\mu=\bar \mu=0.001$. The functions $h_a$, $h_c$ were chosen as
$h_a=h_c=(\sigma(\hat z),\nu \tau,\nu \tau^2)$, $\sigma(\hat z)= 1 / (1 + {\rm exp}(-W_2\hat z+b_2))$, $W_2\in\mathbb{R}^{12\times 7}$, $b_2\in\mathbb{R}^{12}$, 
where $\nu=0.001$. {\color{black}The variations of the terminal costate and its approximated value are displayed in Fig.~\ref{fig:vf_ip}, which shows that the approximated and real terminal costates converge rapidly to the origin, and the approximated terminal costate is close to the desired one.}
\begin{table}[h!tb]
	\centering \caption{Comparisons with r-KMPC in terms of cumulative cost $J$.}
	\label{tab:Tab_com1}
	%
	\renewcommand\arraystretch{1.1}
	\scalebox{0.8}{
		\begin{tabular}{cccccc}
			\toprule 
			Prediction horizon $N$  &8& 12&16& 20&32 \\
			\midrule
			r-LPC (learned online)  &217&218&217&216&217\\
			\midrule
			r-LPC (learned off-line) &149&148&148&147&148\\
			\midrule
			r-KMPC~\cite{XLZ2021} &151&151 &151&151&151\\	
			\bottomrule
		\end{tabular}
	}
\end{table}

\begin{figure}[h]\vspace{0mm}
	\centering
	\includegraphics[width=0.4\textwidth]{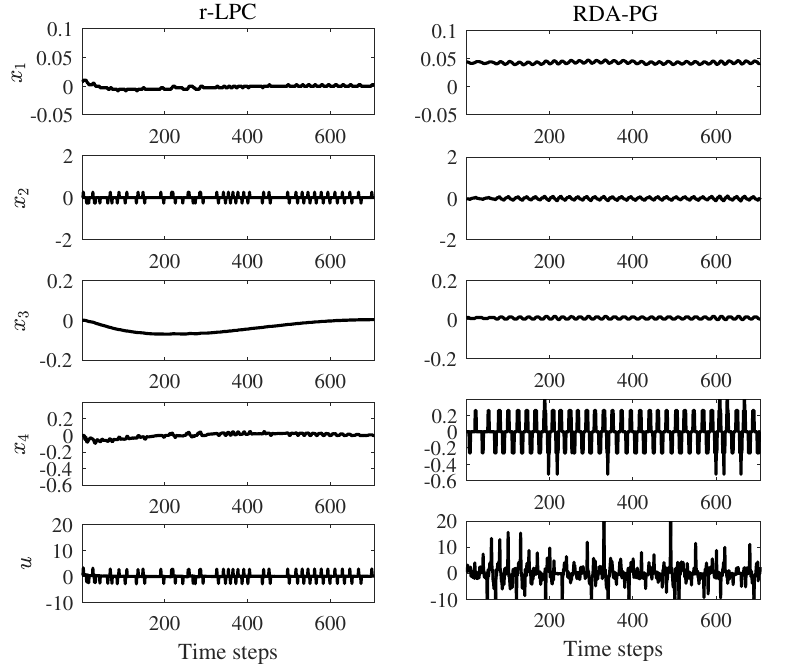}\vspace{0mm}
	\caption{Inverted pendulum: control performance comparison between r-LPC and RDA-DG in the experimental tests under no exogenous disturbance.}\vspace{0mm}
	\label{fig:com-rda-pg}
\end{figure}
\begin{figure}[h]
	\centering
	\includegraphics[width=0.4\textwidth]{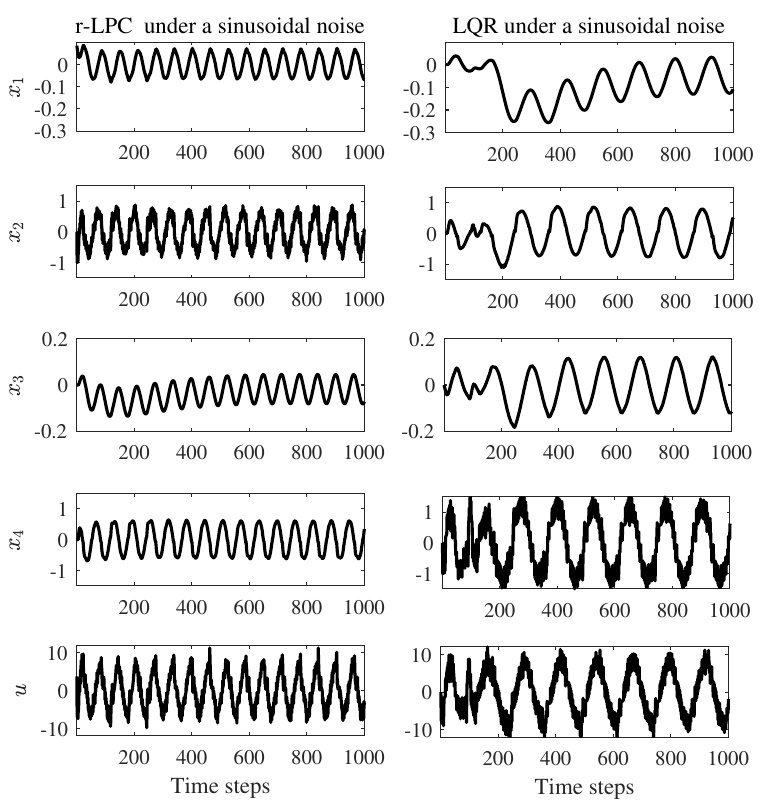}\vspace{0mm}
	\caption{{\color{black}Inverted pendulum: control performance comparison between r-LPC and LQR in the experimental tests under an additive sinusoidal noise, i.e., $w=0.1{\rm sin} (20\pi t)$.}}\vspace{0mm}
	\label{fig:com-lqr}
\end{figure}
\begin{figure}[h]
	\centering
	\includegraphics[width=0.45\textwidth]{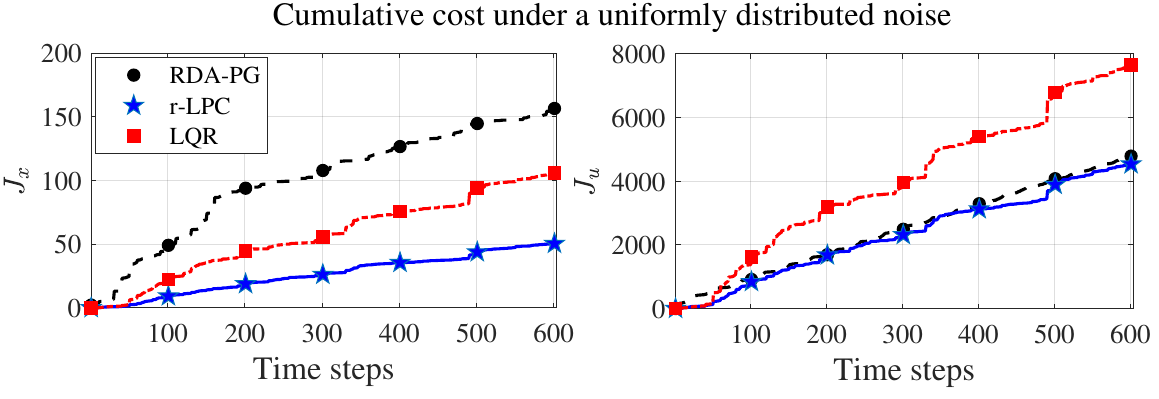}\vspace{0mm}
	\caption{Inverted pendulum: comparison in terms of cumulative cost of the r-LPC, RDA-PG, and LQR in the experimental tests under uniformly distributed noises.}\vspace{0mm}
	\label{fig:com-cost-ip}
\end{figure}
Comparison with r-KMPC in the simulation tests: The r-LPC and r-KMPC were implemented with an initial condition $x(0)=(0.1,0.1,0.2,0.2)$ and a disturbance $w=0.5{\rm sin} (10\pi t)$. The r-LPC was run 10 times to improve the control performance. The final cumulative costs of r-LPC and r-KMPC, i.e., $J=\sum_{j=1}^{N_{\rm sim}}\|x(j)\|_Q^2+\|u(j)\|_R^2$, $N_{\rm sim}=600$,  are collected and displayed in Table~\ref{tab:Tab_com1}, which show that the costs obtained with off-line learned policies are comparable to that of r-KMPC with different prediction horizons.

	\begin{table}[h!tb]\vspace{0mm}
	\centering \caption{Inverted pendulum: comparisons of mean cumulative costs in experimental tests}
	\label{tab:Tab_com}
	\renewcommand\arraystretch{1.3}
	\scalebox{0.75}{
		\begin{tabular}{ccccc}
			\toprule
			\multicolumn{2}{c}{Model}  & Nominal &  S noise&  UD noise\\
			\hline
			{\multirow{2}{*}{r-LPC} }
			&$\bar J_x$&0.01 & $\bm {0.44}$& $\bm {0.08}$ \\ \cline{2-5}
			&$\bar J_u$& 0.79& $\bm {22.3}$ & $\bm {7.51}$\\ \midrule
			{\multirow{2}{*}{RDA-PG~\cite{li2020actor}} }
			&$\bar J_x$&0.03& 10.4& 0.26 \\ \cline{2-5}
			&$\bar J_u$& 12& 31.4& 7.98\\
			\midrule
			{\multirow{2}{*}{LQR} }
			&$\bar J_x$&$\bm {{0.008}}$ & 0.99& 0.17  \\ \cline{2-5}
			&$\bar J_u$&$\bm {0.6}$& ${33.5}$&12.6\\
			\bottomrule\vspace{0mm}
		\end{tabular}
	}
\end{table}
Comparison with an RDA-PG~\cite{li2020actor} and an LQR in the experimental tests: In peculiar, a recently developed regularized policy gradient method, i.e., RDA-PG in~\cite{li2020actor} and  LQR were used for comparisons in the experimental tests.  The control policies used in the experimental studies of r-LPC and RDA-PG were learned off-line. All the experimental tests were performed with the sampling interval $T=0.01$s. The tests were validated in three model conditions, i.e., the inverted pendulum was run under a nominal condition, a sinusoidal (S) noise $w=0.1{\rm sin} (20\pi t)$, and a uniformly distributed (UD) noise with an amplitude of $0.003$ and a frequency of $10$Hz, respectively. The comparative results of r-LPC and RDA-PG in the nominal case are initially displayed in Fig.~\ref{fig:com-rda-pg}, which shows that our approach performs better than RDA-PG. The comparative results of r-LPC and LQR in Fig.~\ref{fig:com-lqr} under the sinusoidal noise illustrate the strong point of our approach to LQR\footnote[1]{A video is also attached as a supplementary document associated with the submission.}. Moreover, the comparative results of the three controllers under the uniformly distributed noise in terms of the cumulative costs $J_x=\sum_{k=1}^{N_{\rm sim}}\|x(k)\|^2$ and $J_u=\sum_{k=1}^{N_{\rm sim}}\|u(k)\|^2$ are displayed in Fig.~\ref{fig:com-cost-ip}, which shows that our approach results in the lowest value of $J_x$ and $J_u$ for the regulation of state and control respectively. For 
comprehensive comparison,  the mean cumulative costs,  $\bar J_x=J_x/N_{\rm sim}$, $\bar J_u=J_u/N_{\rm sim}$ in the prescribed three model scenarios are collected and displayed in Table~\ref{tab:Tab_com}, which shows the effectiveness of r-LPC in control of systems under disturbances when compared with RDA-PG and LQR.
\section{Conclusions}

This paper proposes a robust learning-based predictive control scheme, i.e., r-LPC, for nonlinear discrete-time dynamical systems with unknown dynamics, state constraints, and uncertainties. In r-LPC, instead of numerically computing a control sequence by a nominal MPC,  an actor-critic learning algorithm is proposed to learn an explicit time-dependent control policy in a receding-horizon manner. The resulting output of r-LPC is composed of an RL-based control policy and an off-line nonlinear state-feedback policy. As the prominent feature,  r-LPC can guarantee feasibility in each prediction interval if state constraint violations occur by actor-critic learning, which allows applying our approach to control safety-critical systems. The closed-loop robustness with r-LPC under approximation errors is proven, and asymptotic stability under no exogenous disturbance is obtained. 

The proposed r-LPC algorithm has advantages over previous robust MPC and model-based optimal control methods. For instance, compared with tube MPC, see~\cite{XLZ2021,falugi2013getting},  r-LPC results in an explicit control policy, allowing for control implementation with limited computational resources; while compared to LQR, the advantage lies in the resultant nonlinear control policy,  the constraint fulfilment, and the online learning ability due to the actor-critic learning.  Future works will focus on the extension of r-LPC to continuous-time nonlinear systems.

\vspace{0mm}
\appendices
\section{}\label{appen-a}
\subsection{Proof of Theorem~\ref{the:ac-conver}}\label{sec:proof_2} 
1) Consider the Lyapunov function
$$
\begin{array}{ll}
L(\tau)=L_a(\tau)+L_c(\tau)\\
\end{array}
$$
where $L_{\star}=\frac{1}{2}\text{tr}(\tilde W_{\star}^{\top}\tilde W_{\star})$, $\tilde W_{\star}=W_{\star}-\hat W_{\star}$, $\star=a,c$ in turns. In view of~\eqref{Eqn:wc} and~\eqref{Eqn:wa}, the difference of $L$ writes
\begin{equation}\label{Eqn:deltaL}
\Delta L(\tau+1)=\Delta L_a(\tau)+\Delta L_c(\tau)
\end{equation}
where

\begin{small}
\begin{equation}\label{Eqn:deltaL_c}	
\begin{array}{llll}
	\Delta L_{\star}(\tau)=\underbrace{\text{tr}(\gamma_{\star}\nabla\delta_{\star}(\tau)^{\top}\tilde  W_{\star}(\tau))}+\underbrace{0.5\text{tr}({\gamma_{\star}}^2\nabla\delta_{\star}(\tau)^{\top}
\nabla\delta_{\star}(\tau))}\\
	\hspace{27mm}\nabla L_{\star,1}(\tau) \hspace{22mm} \nabla L_{\star,2}(\tau)
	\end{array}	
\end{equation}
\end{small}$\nabla\delta_{\star}=\partial\delta_{\star}/\partial \hat W_{\star}$, $\star=a,c$ in turns. 
To compute $\Delta L_c$, in view of~\eqref{Eqn:ec}, one writes 
\begin{small}
	\begin{equation}\label{Eqn:partial error c}
	\begin{array}{llll}
\nabla\delta_{c}(\tau)=\underbrace{\frac{\partial \text{tr}(\|\epsilon_{c}(\tau)\|^2+\|\epsilon_{c,N}\|^2)}{\partial \hat W_{c}}}+\underbrace{\frac{\partial\text{tr}(\bar\mu (\mathcal{B}(\hat\lambda(\tau))+\mathcal{B}_f(\hat\lambda)))}{\partial \hat W_{c}}}\\
\hspace{27mm}\nabla\delta_{c,1}(\tau) \hspace{25mm} \nabla\delta_{c,2}(\tau)
\end{array}
\end{equation}
\end{small}%
For the sake of simplicity, in the rest of the Appendix we use $q$ to denote a generic variable $q$ with time index $\tau$, i.e., $q(\tau)$, $q^{+}$ to represent $q(\tau+1)$, $q_N=q(k+N)$. 
In view of the definition of $\lambda_d$ and $\lambda^{\ast}$, it holds that
\begin{equation}\label{Eqn:error}
\begin{array}{ll}
\epsilon_{c}\hspace{0mm}&=\lambda_d-\lambda^{\ast}+\lambda^{\ast}-\hat \lambda\\
&=\xi_c+\Delta \bar \epsilon_c
\end{array}
\end{equation}
where $\xi_c=-A^{\top}\tilde W_c^{\top}h_c^++\tilde W_c^{\top}h_c$, $\Delta \bar \epsilon_c=\bar \epsilon_c-A^{\top}\bar \epsilon_c^+$. 
The term $\nabla\delta_{c,1}$ in~\eqref{Eqn:partial error c} can be computed as
\begin{equation}\label{Eqn:Wc-fac}\begin{array}{lll}
\nabla\delta_{c,1}
&=&-2(h_c(\xi_c+\Delta\bar\epsilon_c)^{\top}-h_c^{+}(\xi_c+\Delta\bar\epsilon_c)^{\top}A^{\top}\\
&&+h_{cN}(\xi_{cN}+\bar\epsilon_{cN})^{\top})
\end{array}
\end{equation}
where $\xi_{cN}=\tilde{W}_c^{\top}h_{cN}$, $h_{cN}=h_c(k+N)$. 

Consider the constraint on $\hat \lambda$ of type ${{\Lambda}}^{i}=\{\hat \lambda|\hat \lambda^{\top}Z_{i}\hat\lambda\leq 1\}$, then one can compute the term $\nabla\delta_{c,2}$ in~\eqref{Eqn:partial error c} as

\begin{equation}\label{Eqn:Wc-bar-fac}\begin{array}{ll}
\nabla\delta_{c,2}
	=2\bar\mu (\kappa_{\lambda}^{-1}h_ch_c^{\top}\hat W_cZ_{i}+\kappa_{\lambda N}^{-1}h_{cN}h_{cN}^{\top}\hat W_cZ_{N}):=2\bar\mu\mathcal{\omega}
\end{array}
\end{equation}
where 
 $\kappa_{\lambda}=1-\hat \lambda^{\top}Z_{i}\hat\lambda$,
 $\kappa_{\lambda N}=1-\hat \lambda^{\top}Z_{N}\hat\lambda$.
 
In view of~\eqref{Eqn:Wc-fac} and \eqref{Eqn:Wc-bar-fac}, one can write the first term in~\eqref{Eqn:deltaL_c} as
\begin{equation}\label{Eqn:DELTA LC}
\begin{array}{ll}
\nabla L_{c,1}\hspace{0mm}&=-2\gamma_c \text{tr}(\xi_c(\xi_c+\Delta\bar \epsilon_c)^{\top}\hspace{-1mm}+\hspace{-1mm}\xi_{cN}(\xi_{cN}+\bar \epsilon_{cN})^{\top}\hspace{-1mm}-\hspace{-1mm}\bar\mu\tilde{W}_c^{\top}\omega)\\
\hspace{0mm}&=-\text{tr}(\|\bar \xi_c\|^2_{M}+\bar \xi_c^{\top}M\rho_1-2\gamma_c\bar\mu\tilde{W}_c^{\top}\omega)
\end{array}
\end{equation}
where $\bar \xi_c=(
\xi_c,\xi_{cN})$, $\rho_1=(
\Delta\bar \epsilon_c,\bar \epsilon_{cN})$, $M=2\gamma_cI_2$.
Also, one can compute:  
\begin{equation}\label{Eqn:DELTA LC1}
\begin{array}{ll}
\nabla L_{c,2}=\text{tr}(-\|\bar \xi_c\|_{G_1}^2+\bar \xi_c^{\top}(G_2\rho_1+G_3\rho_2)+\bar g_{\bar \epsilon_c}
)\end{array}
\end{equation}
where  $\rho_2=
(\hat W_c^{\top}h_c,\hat W_c^{\top}h_{cN})
$, 
$\bar g_{\bar \epsilon_c}=2\gamma_c^2(\varphi_c^{\top}\varphi_c+2\bar\mu\varphi_c^{\top}\omega+\bar\mu^2\omega^{\top}\omega))$,  $\varphi_c=-h_c\Delta\bar\epsilon_c^{\top}+h_c^{+}\Delta\bar\epsilon_c^{\top}A^{\top}-h_{cN}\bar\epsilon_{cN}^{\top}$, $$ G_{1}=-\begin{bmatrix}
g_1&g_{12}^{\top}\\g_{12}& g_2
\end{bmatrix},$$ 
 $g_1=2\gamma_c^{2}(\bar h_c+\bar h_c^{+}A^{\top}A-2A^{\top}h_c^{\top}h_c^{+})$, $g_{2}=2\gamma_c^{2}\bar h_{cN}$, $g_{12}=2\gamma_c^{2}(h_c^{\top}h_{cN}-A(h_c^{+})^{\top}h_{cN})$, $\bar q=q^{\top}q$ for $q=h_c,h_{cN},h_c^{+}$,
$$G_{2}=\begin{bmatrix}
	g_3&g_{4}\\g_5&g_{6}
\end{bmatrix},$$
and where $g_3=4\gamma_c^2(\bar h_c-h_c^{\top}h_c^{+}A-(h_c^{+})^{\top}h_cA^{\top}+\bar h_c^{+}AA^{\top})$, $g_4=4\gamma_c^2(-(h_c^{+})^{\top}h_{cN}A+h_c^{\top}h_{cN})$,   $g_5=4\gamma_c^2(h_{cN}^{\top}h_c-h_{cN}^{\top}h_c^{+}A)$, $g_6=4\gamma_c^2h_c^{\top}h_{cN}$,
$$G_{3}=\begin{bmatrix}
g_{7}& g_8
\\g_{9}& 
g_{10}
\end{bmatrix},$$
and where $g_7=4\bar\mu\gamma_c^2\kappa_{\lambda}^{-1}(-\bar h_c+(h_c^{+})^{\top}h_cA)Z_i$,
$g_8=4\bar\mu\gamma_c^2\kappa_{\lambda N}^{-1}(-h_c^{\top}h_{cN}+(h_c^{+})^{\top}h_{cN})A)Z_N$,
$g_9=-4\bar\mu\gamma_c^2\kappa_{\lambda}^{-1}h_{cN}^{\top}h_cZ_i$, $g_{10}=-4\bar\mu\gamma_c^2\kappa_{\lambda N}^{-1}\bar h_{cN}Z_N$.

Taking~\eqref{Eqn:DELTA LC} and~\eqref{Eqn:DELTA LC1} into consideration, $\Delta L_c$ results
\begin{equation}\label{Eqn:delta Lc-expan}
\begin{array}{ll}
\Delta L_c=\text{tr}(-\|\bar \xi_c\|_{G_{c1}}^2+\bar \xi_c^{\top}(G_{c2}\rho_1+G_{c3}\rho_2)+ g_{\bar \epsilon_c}
)\end{array}
\end{equation}
where $G_{c1}=G_1+M$, $ G_{c2}=G_2-M$, $G_{c3}=G_3$, $g_{\bar \epsilon_c}=\bar g_{\bar \epsilon_c}+2\gamma_c\bar\mu\tilde{W}_c^{\top}\omega$.

Applying the Young's inequality and the Cauthy–Schwartz inequality, it holds that
$$\begin{array}{ll}
\rm{tr}(\bar \xi_c^{\top}(G_{c2}\rho_1+G_{c3}\rho_2))\leq&
\frac{\alpha_c}{2}(\|G_{c2}\|^2+\|G_{c3}\|^2)\|\bar \xi_c\|^2+\\
&\frac{1}{2\alpha_c}(\|\rho_1\|^2+\|\rho_2\|^2)
\end{array}$$
where $\alpha_c>0$.
{\color{black}
As $\hat W_c$ is bounded provided $\delta_c$ is finite, we assume $\|\hat {W}_c\|\leq W_{c,m}$, then $\|\tilde {W}_c\|\leq 2W_{c,m}$ in view of Assumption~(A4). One can promptly has $\|\omega\|\leq\bar \omega$, $\|\varphi_c\|\leq\bar \varphi_c$, $\| g_{\bar \epsilon_c}\|\leq g_{c,m}$, where the derivation of $\bar\omega$, $\bar\varphi_c$, and $g_{c,m}$ is trivial and neglected for simplicity.
}
Hence,~\eqref{Eqn:delta Lc-expan} leads to
\begin{equation}\label{Eqn:delta Lc-bound}
\begin{array}{ll}
\Delta L_c(\tau)\leq -\|\bar \xi_c(\tau)\|_{\bar G_{c1}}^2+{\rm res_c}
\end{array}
\end{equation}
where $\bar G_{c1}=G_{c1}+\frac{\alpha_c}{2}(\|G_{c2}\|^2+\|G_{c3}\|^2)$, ${\rm res_c}= g_{c,m}+\frac{1}{2\alpha_c}(2\bar\epsilon_{c,m}^2+2W_{c,m}^2h_{c,m}^2)$.


As a second step, we compute the $\Delta L_a$ in~\eqref{Eqn:deltaL}.
Note that, in view of~\eqref{Eqn:loss_a}, one has $$
\nabla\delta_{a}=\frac{\partial \text{tr}(\|\epsilon_{a}\|^2+\bar\mu \mathcal{B}(\hat u))}{\partial \hat W_{a}}$$
Consider the input constraint of type $\hat{\mathcal{U}}=\{\hat u|\sum_{i=1}^p a_{u,i}^\top \hat u\leq 1,\forall i\in\mathbb{R}_{1}^p\}$. 
	One can write $\frac{\partial B(\hat u)}{\partial \hat u}=\sum_{i=1}^pa_{u,i}^{\top}\kappa_{\hat u,i}^{-1}$, where $\kappa_{\hat u,i}=1-a_{u,i}^{\top}\hat u$. Hence,
\begin{equation}\label{Eqn:Wa-fac}
\begin{array}{ll}
\frac{\partial \text{tr}(\|\epsilon_{a}\|^2)}{\partial \hat W_{a}}
=-2h_a(\varphi_a+\bar R_2\tilde{W}_a^{\top}h_a)^{\top}\bar R_1
\end{array}
\end{equation}
where $\varphi_a=(\tilde a_{u,i}+I)\bar\epsilon_{a}+B^{\top}\tilde W_c^{\top}h_c^{+}+B^{\top}\bar\epsilon_c^{+}$, $ \bar R_1=2R+\kappa_{\hat u,i}^{-2}\tilde a_{u,i}$, $\bar R_2=2R+\sum_{i=1}^p\tilde a_{u,i}\kappa_{u,i}^{-1} \kappa_{\hat u,i}^{-1}$, $\kappa_{u,i}=1-a_{u,i}^{\top} \hat u_d$, $\tilde a_{u,i}=a_{u,i}a_{u,i}^{\top}$.
Recalling the fact that $\frac{\partial a^{\top }X^{\top}b}{\partial X}=ba^{\top}$ where $X$ is a matrix, $a$ and $b$ are two vectors, one has
\begin{equation}\label{Eqn:Wa-bar-fac}\begin{array}{ll}
\frac{\partial \text{tr}( \mathcal{B}(\hat u(\tau)))}{\partial \hat W_{a}(\tau)}= h_al_{u}
\end{array}
\end{equation}
where $l_{u}=\sum_{i=1}^p{a_{u,i}^{\top}}{\hat\kappa_{u,i}}^{-1}$.
Hence, denoting $\xi_a=\tilde{W}_a^{\top}h_a$, in view of \eqref{Eqn:Wa-fac} and~\eqref{Eqn:Wa-bar-fac}, it follows that
$$\begin{array}{ll}
\Delta L_a=\text{tr}(-\|\xi_a\|_{G_{a1}}^2+G_{a2}\xi_a+g_{\bar \epsilon_a})\\
\end{array}$$
where $G_{a1}=(2\gamma_a I-2\gamma_a^2\bar h_a\bar R_2\bar R_1)\bar R_1\bar R_2$, $G_{a2}=(4\gamma_a^2\bar R_2\bar R_1\bar R_1\varphi_a\bar h_a-2\gamma_a\bar R_1\varphi_a+\bar\mu\gamma_a lu^{\top}+2\gamma_a^2\bar\mu\bar R_2\bar R_1 lu^{\top}\bar h_a)^{\top}$, $\bar h_a=h_a^{\top}h_a$,
$g_{\bar \epsilon_a}=2\gamma_a^2\bar R_1\varphi_a \bar h_a\varphi_a^{\top}\bar R_1-2\gamma_a^2\bar\mu lu^{\top}\bar h_a\phi_{a}\bar R_1+1/2\gamma_a^2\bar\mu^2lu^{\top}\bar h_alu$,

Likewise, in view of Assumption~(A4), one can promptly has $\|\varphi_a\|\leq\bar {\phi}_a$, $\|l_{u}\|\leq\bar l_u$, $\|g_{\bar \epsilon_a}\|\leq g_{a,m}$, and  $\|G_{a2}\|\leq G_{am}$, where the derivation of $\bar {\phi}_a$, $\bar l_u$, $g_{a,m}$, and $G_{am}$ is trivial and is neglected for simplicity.
Applying the Young's inequality and the Cauthy–Schwartz inequality leads to
\begin{equation}\label{Eqn:delta La-bound}
\begin{array}{ll}
\Delta L_a
\leq -\|\xi_a\|_{\bar G_{a1}}^2+{\rm res_a}
\end{array}
\end{equation}
where $\bar G_{a1}= G_{a1}-\frac{\alpha_a\gamma_a}{2}$, $\alpha_a>0$ is a tuning parameter,  ${\rm res_a}=\frac{2}{\alpha_a\gamma_a}G_{am}^2+ g_{a,m}$. 
With~\eqref{Eqn:delta Lc-bound} and~\eqref{Eqn:delta La-bound}, one can write
\begin{equation}
\Delta L\leq -\|\bar \xi_c\|_{\bar G_{c1}}^2-\|\xi_a\|_{\bar G_{a1}}^2+{\rm res}
\end{equation}
where ${\rm res}={\rm res_{a}}+{\rm res_{c}}$, leading to the variables $\bar \xi_c(\tau)$ and $\xi_a(\tau)$ converging to the corresponding sets as $\tau\rightarrow N$ and $N\rightarrow +\infty$, i.e.,
\begin{equation}\label{Eqn:uub}
\begin{array}{ll}
\|\bar \xi_c(\tau)\|\leq \frac{\rm res}{\sigma_{\rm min}(\bar G_{c1})},\
\| \xi_a(\tau)\|\leq \frac{\rm res}{\sigma_{\rm min}(\bar G_{a1})}.
\end{array}
\end{equation}
In view of the definition of $\bar \xi_c$ (cf.~\eqref{Eqn:error}) and $ \xi_a$, one has  $$\begin{array}{ll}
\bar \xi_c=\epsilon_c^{l}-\tilde \epsilon_{c},\
 \xi_a=\epsilon_a-\bar \epsilon_{a}
\end{array}$$
where $\epsilon_c^{l}=(\epsilon_c,\epsilon_{cN})$, $\tilde \epsilon_{{c}}=(\Delta \bar \epsilon_c,\bar \epsilon_{cN})$.
Hence,~\eqref{Eqn:uub} results
\begin{subequations}\label{Eqn:est-uub}
\begin{align}
\|\epsilon_c^{l}(\tau)\|\leq \frac{\rm res}{\sigma_{\rm min}(\bar G_{c1})}+\pi_{A} \epsilon_{c,m}:=\eta_c,\label{Eqn:epi_ct}\\
\| \epsilon_a(\tau)\|\leq \frac{\rm res}{\sigma_{\rm min}(\bar G_{a1})}+\bar \epsilon_{a,m}:=\eta_a.
\end{align}
\end{subequations}
where the inequality in~\eqref{Eqn:epi_ct} comes from that $\|\tilde \epsilon_{{c}}\|=\sqrt{\Delta \bar \epsilon_c^{\top}\Delta \bar \epsilon_c+\bar \epsilon_{cN}^{\top}\bar \epsilon_{cN}}\leq\pi_{A} \epsilon_{c,m} $, $\pi_{A}=\sqrt{2+2\|A\|+\|A^2\|}$.

2) Provided that $\bar \epsilon_{c}=0$, $\bar \epsilon_{{cN}}=0$, $\bar\mu=0$, one has
$\Delta L_c(\tau)\leq -\|\bar \xi_c(\tau)\|_{\bar G_{c1}}^2$ 
which implies that $\bar \xi_c(\tau)\rightarrow 0$ exponentially. That is to say, $\hat\lambda(\tau)$ converges to $\lambda_d(\tau)$ at an exponential rate. In view of this, one can rewrite~\eqref{Eqn:lam_d} as
\begin{equation}\label{}
\begin{array}{l}
\lambda_d^{i+1}(\hat {z}(\tau))=
\mu \frac{\partial \mathcal{B}(\hat {z}(\tau))}{\partial \hat {z}(\tau)}+
2\bar Q\hat {z}(\tau)+
A^\top \lambda_d^i(\hat {z}(\tau+1))
\end{array}
\end{equation}
where the superscript $i$ is the iterative step corresponding to $\hat W_c(\tau+i)$, i.e., $\lambda_d^{i}(\hat {z}(\tau))=\hat W_c(\tau+i)^{\top}h_c(\tau)$.
In view of~\eqref{Eqn:lambda} and denoting $\tilde \lambda$ as the subtraction of  $\lambda^{\ast}$ and $\lambda_d$, one promptly has
$\tilde \lambda^{i+1}(\tau)=A^{\top}\tilde \lambda^{i}(\tau+1)$.
By induction, one has
$$\tilde \lambda^{i}(\tau)=(A^{\top})^i\tilde \lambda^0(\tau+i)\rightarrow 0$$
as $i\rightarrow+\infty$, due to the fact $\|A^{\top}\|\rightarrow 0$ for $i\rightarrow +\infty$ in view of $A$ being Schur stable.
This also implies $\hat \lambda(\tau)\rightarrow\lambda^{\ast}(\tau)$ for $\tau\in[k,k+N-1]$ as $N\rightarrow +\infty$. With this result, the term $\hat W_c^{\top}h_c^{+}$ in~\eqref{Eqn:Wa-fac} is a vanishing term, then one has 
\begin{equation*}
\begin{array}{ll}
\Delta L_a
\leq -\|\xi_a\|_{\bar G_{a1}}^2,
\end{array}
\end{equation*}
leading to the result  that $\xi_a$ converges to $\xi_a^{\ast}$. Hence, $\hat W_c(\tau)\rightarrow W_c$ and $\hat W_a(\tau)\rightarrow W_a$  as $\tau\rightarrow N$ and $N\rightarrow+\infty$.
 \hfill$\square$\vspace{-1.5mm}
\subsection{Proof of Theorem~\ref{the:2}}\label{appen:the2}
 Assume that at any time instant $k$,
a feasible control sequence can be generated with~\eqref{Eqn:actor} at time $k$. We denote the near-optimal control policy as 
$\hat u(\hat z(k|k)),\cdots,\hat u(\hat z(k+N-1|k))$ 
associated with a near-optimal cost given as $\bar V(k)$ such that $x(k+i|k)\in\mathcal{X}$, $u(k+i|k)\in \mathcal{U}$, $\forall i\in[0,N-1]$, $\hat{z}(k+N|k)\in\mathcal{Z}_f$. At the next time instant $k+1$,~\eqref{Eqn:safe-policy} is a feasible choice such that $x(k+i|k+1)\in\mathcal{X}$, $u(k+i|k+1)\in \mathcal{U}$, $\forall i\in[1,N]$, $\hat{z}(k+N+1|k+1)\in\mathcal{Z}_f$. This is due to the standard recursive feasibility argument of MPC, see~\cite{mayne2005robust}.  Hence, with~\eqref{Eqn:safe-policy}, the recursive feasibility of the finite-horizon RL follows. 
Moreover, provided the feasible policy $\bm {\hat u}^s(\hat z(k+1))$ (see~\eqref{Eqn:safe-policy}) at time $k+1$, the corresponding weight satisfies  
$\hat W_a^{\top}{\bm h}_a=\bm {\hat u}^s(\hat z(k+1))$
leading to~\eqref{Eqn:safe-weight}  in view of~\eqref{Eqn:persistent}. \hfill$\square$
\vspace{-1.5mm}
\subsection{Proof of Theorem~\ref{the:3}}\label{appen:the3}
 At time instant $k+1$, the learned cost $\bar V(k+1|k+1)$ might not be smaller than  $\bar V(k+1|k)$. In this case and the case that feasibility is not guaranteed, the control policy~\eqref{Eqn:safe-u} can be applied.
Hence, denoting $\bar V^{s}(k+1)=\min\{\bar V(k+1|k+1),\bar V(k+1|k)\}$,  one has
\begin{equation}\label{Eqn:V-MONO}
\begin{array}{ll}
\bar V^{s}(k+1)- \bar V^{s}(k)\leq
-(\|\hat {z}(k)\|_{\bar Q}^2+\|\hat u(k)\|_{R}^2
+\mu \mathcal{B}(\hat z(k))+\\\mu \mathcal{B}(\hat u(k)))-\mu \mathcal{B}_f(\hat {z}(k+N))
+\mu \mathcal{B}_f(\hat {z}(k+N+1))+\\\|\hat {z}(k+N)\|_{F^{\top}PF-P+\bar Q+K^{\top}RK+\mu H}
\end{array}
\end{equation}
Recalling that
\begin{equation}
\begin{array}{ll}
\mathcal{B}_f(\hat {z}(k+N+1))- \mathcal{B}_f(\hat {z}(k+N))\\
=\frac{1}{1-{\hat {z}(k+N)^{\top}}F^{\top}ZF\hat {z}(k+N)}-\frac{1}{1-{\hat {z}(k+N)^{\top}}Z\hat {z}(k+N)}\\
<0,
\end{array}
\end{equation}
in view of~\eqref{Eqn:qn_ph}, from~\eqref{Eqn:V-MONO}, the monotonic property of the value function is  obtained, i.e., $
\bar V(k+1)- \bar V(k)\leq
-(\|\hat {z}(k)\|_{\bar Q}^2+\|\hat u(k)\|_{R}^2
+\mu \mathcal{B}(\hat z(k))+\mu \mathcal{B}(\hat u(k))) $
which leads to $\bar V(k+1)- \bar V(k)\rightarrow0$ as, $k\rightarrow\infty$. Hence $\hat {z}(k)$, $\hat u(k)\rightarrow0$ asymptotically. Consequently $\hat x(k)\rightarrow 0$ asympotically since $\hat x=C\hat z$. 
Recall that the real state remains inside the tube, i.e., $\hat z\oplus{\mathcal{Z}}\in\mathcal{X}$, then the robustness is obtained, i.e.,
${z}(k)\rightarrow\mathcal{O}_z$, consequently
$x(k)\rightarrow {\mathcal{O}}_x$ asymptotically.
\hfill$\square$
\vspace{-1.5mm}
\subsection{Proof of Theorem~\ref{the:rmpc-convergence}}\label{appen:the:rmpc-convergence}
We present a different proof to~\cite{XLZ2021}. Recalling that $\hat u$, $\hat z$, $\hat x\rightarrow 0$ (cf. Theorem~\ref{the:3}) and $w=0$, one can write~\eqref{Eqn:linear_p-residual} as
\begin{equation}\label{Eqn:linear_p-residual-no-dis}
\left\{\begin{array}{ll}
{z}(k+1)=F{z}(k)+\bar w(k),\|\bar w(k)\|\leq  \|z(k)\|_{\bar L}\\
x(k)=C{z}(k)+v(k)
\end{array}\right.
\end{equation}
Consider the Lyapunov function $V(k)=z(k)^{\top}Pz(k)$. To guarantee asymptotic stability, its difference $\Delta V(k)=z(k+1)^{\top}Pz(k+1)-z(k)^{\top}Pz(k)=z(k)^{\top}(F^{\top}PF-P)z(k)+2z(k)^{\top}F^{\top}Pw(k)+w(k)^{\top}w(k)<0$, leading to the linear matrix inequality~\eqref{Eqn:iss-con} by applying the S-procedure in~\cite{1994Linear} with the condition $\|\bar w(k)\|\leq  \|z(k)\|_{\bar L}$.
Indeed, with~\eqref{Eqn:iss-con}, one has that $z(k)\rightarrow 0$ asymptotically. Consequently, it follows that  $u(k)\rightarrow 0$ asymptotically. In view of $\Psi(0)=0$ one has $v=0$ as $x=0$. Hence, $x=Cz+v\rightarrow 0$ asymptotically. \hfill$\square$
\vspace{-1.5mm}
\subsection{Derivation of $A$, $B$, $C$, $D$ and sets $\mathcal{D}$, $\mathcal{V}$}\label{SEC:COM-SETS}
Inline with~\cite{korda2018linear,XLZ2021}, the matrices $A$, $B$, $C$, and $D$ can be computed using input and output datasets. Let us assume to have $L$ datasets of $(w_{i},u_i,x_i,x_i^+)$, where $x_i^+$ is the successor state of $x_i$ by applying $(w_{i},u_i)$ to~\eqref{Eqn:non-model}, $w_{i}$ can be computed using nonlinear estimation methods if it is not measurable, see~\cite{korda2018linear}. It is assumed that $\{(u_i,x_i)\}_{i=1}^L$ are drawn independently according to a non-negative probability distribution. Then, $A$, $B$, $C$, and $D$ can be obtained via minimizing a type of cost function defined as
\begin{equation}\label{Eqn:appro_K}
\begin{array}{ll}
{\rm Loss}=&\sum_{i=1}^{L}\|A\Psi(x_i)+Bu_i+Dw_i-\Psi(x_i^+)\|^2+\vspace{1mm}\\
&\|C{\Psi(x_i)}-x_i\|^2+\alpha\|[A\ B\ D]\|^2+\beta\|C\|^2
\end{array}
\end{equation}
where $\alpha$ and $\beta$ are positive scalars. With~\eqref{Eqn:appro_K} being solved, sets $\mathcal{D}$ and $\mathcal{V}$ can be estimated according to the empirical risk evaluation in \cite{ hertneck2018learning}. For computational details on the derivation of $\mathcal{D}$ and $\mathcal{V}$ please refer to~\cite{XLZ2021}.

\bibliographystyle{IEEEtran}
\bibliography{IEEEabrv,ref}
\end{document}